\documentclass[pra,twocolumn,showpacs,footinbib]{revtex4}%
\usepackage{amsfonts}
\usepackage{graphicx}
\usepackage{amsmath}
\usepackage{amssymb}
\usepackage{amscd}%
\begin{document}
\title{Theory of decoherence in a matter wave Talbot-Lau interferometer}
\author{Klaus Hornberger}
\altaffiliation[present address: ]{Department f\"{u}r Physik, Ludwig-Maximilians-Universit{\"{a}}t,
Theresienstra{\ss }e 37, 80333 M{\"{u}}nchen, Germany}

\author{John E. Sipe}
\altaffiliation[permanent address: ]{Department of Physics, University of Toronto, 60 St.~George Street, Toronto,
ON, Canada M5S 1A7}

\author{Markus Arndt}
\affiliation{Institut f{\"u}r Experimentalphysik, Universit{\"a}t Wien, Boltzmanngasse 5,
1090 Wien, Austria}

\begin{abstract}
We present a theoretical framework to describe the effects of decoherence on
matter waves in Talbot-Lau interferometry. Using a Wigner description of the
stationary beam the loss of interference contrast can be calculated in closed
form. The formulation includes both the decohering coupling to the environment
and the coherent interaction with the grating walls. It facilitates the
quantitative distinction of genuine quantum interference from the expectations
of classical mechanics. We provide realistic microscopic descriptions of the
experimentally relevant interactions in terms of the bulk properties of the
particles and show that the treatment is equivalent to solving the
corresponding master equation in paraxial approximation.

\end{abstract}

\date{July 29,2004 }
\maketitle


\section{Introduction}

The art of demonstrating the wave nature of material particles experienced
considerable advances in recent years; see
\cite{Berman1997a,Rauch2000a,Martellucci2000a} and references therein. The
interfering species evolved from the elementary particles of the early
experiments \cite{Davisson1927a,Halban1936a} to composite objects with an
internal structure. In particular, the experiments in atom interferometry have
left the stage of proof-of-principle demonstrations, and provide substantial
applications in metrology \cite{Weiss1993a,
Ekstrom1995a,Gustavson1997a,Peters1999a}. Objects with even larger complexity,
such as molecules or clusters, exhibit a rich internal structure that can
interact in various ways with external fields. Their interference is highly
sensitive to the corresponding phase shifts, thus offering the potential to
measure molecular properties with unprecedented precision. At the same time
any coupling to uncontrollable fields and environmental degrees of freedom
severely limits the ability of large objects to show interference. These
effects are bound to become dominant as the chosen objects increase in size
and complexity.

The influence of environmental coupling on a quantum system may be described
by decoherence theory \cite{Joos2003a,Zurek2003a}. It considers both the
influence of noise due to uncontrollable external fields and the effect of the
entanglement with unobserved dynamic degrees of freedom. This latter
phenomenon -- the dynamic delocalization of quantum coherence into many
environmental degrees of freedom -- largely explains the emergence of
classical behavior in a quantum description. In particular, it describes the
wave-particle complementarity encountered if one seeks to determine by a
(macroscopic) measurement device the path taken. Since matter wave
interferometers establish quantum coherence on a macroscopic scale they are
sensitive tools to probe the quantum-to-classical transition of complex objects.

The purpose of this article is to provide the theoretical framework needed to
describe the diffraction and decoherence effects encountered in the
interferometry of large, massive objects. We focus on near-field Talbot-Lau
interference, which is the favored setup for short de Broglie wavelengths. We
take care to describe the effects of diffraction and decoherence with
realistic parameters, to permit a direct quantitative comparison with the
experimental signal. The interactions are treated on a microscopic level using
the bulk properties of materials and particles. We note that the recent
interference experiments with fullerenes and biomolecules
\cite{Brezger2002a,Hornberger2003a,Hackermuller2003a,Hackermuller2004a} were
analyzed using the theory presented in this article.

Before going into calculations we start with an informal discussion of
Talbot-Lau interference \cite{Clauser1994a,Chapman1995c,Nowak1997a}. In this
setup an essentially uncollimated particle beam passes three parallel
gratings. Effectively, the first grating acts as an array of collimation slits
which illuminate the second grating. Diffraction at the second grating then
leads, for particular choices of the grating periods and the wavelength, to a
high contrast near-field interference pattern at the position of the third
grating. This density pattern is observed with the help of the third grating
by recording the transmitted flux as a function of the lateral grating position.

An important advantage of the Talbot-Lau effect is the favorable scaling
behavior with respect to larger masses of the interfering object
\cite{Clauser1997a}. Unlike in far-field diffraction, where the required
grating period falls linearly with the de~Broglie wavelength, it decreases
merely like the square root in the Talbot-Lau setup. In addition, the
collimation requirements are much weaker than for far field diffraction, and
the spatially resolving detector is already built into the device.

However, for a fixed particle velocity it is not immediately evident whether
the observed signal proves genuine quantum interference, since a certain
fringe pattern could also be expected from a classical moir\'{e} effect. This
classical pattern can be suppressed by an appropriate choice of the open
fraction of the grating and, unlike the strong wave-length dependence found in
the interference effect, the ideal classical shadow fringes do not depend on
the velocity. Nonetheless, in order to distinguish clearly the quantum
phenomenon from a classical expectation it is necessary to describe the
quantum and the classical evolution in the same theoretical framework, thus
ensuring that all interactions and approximations are treated equally.

A first aim of this article is to provide such a description that draws the
demarcation line between the predictions of quantum and classical mechanics
concisely and quantitatively. The second aim is then to account for the
relevant environmental interactions, thus providing a quantitative description
of the transition from the quantum to the classical behavior. For both goals
it will be helpful to describe the state in the interferometer in terms of a
stationary, unnormalized Wigner function. Due to the stationary formulation
the effect of decoherence will not be given by a master equation. Therfore, it
is shown in the final part of the paper that our treatment is equivalent to
the conventional dynamic formulation of decoherence in terms of a normalized
Wigner function.

The structure of the article is as follows: In Sect.~\ref{sec:wigner} we
review the coherent Talbot-Lau effect and give a formulation in terms of the
Wigner function. The corresponding classical shadow effect is calculated on an
equal footing in the phase space representation. The influence of the
interaction with realistic gratings, which is very important for a
quantitative description, is accounted for in Sect.~\ref{sec:gratings}. Those
effects are also treated on an equal degree of approximation in the quantum
and the classical description. In Sect.~\ref{sec:deco} we include the
possibility of decoherence and show how it can be accounted for analytically.
The specific predictions for decoherence due to collisions and due to heat
radiation are then obtained in Sect.~\ref{sec:decofunction}. In
Sect.~\ref{sec:mastereq} we relate the description of decoherence in terms of
a stationary beam to the solution of the corresponding time-dependent master
equation. Concluding remarks are given in Sect.~\ref{sec:conclusions}.

\section{The Talbot-Lau effect in the Wigner representation}

\label{sec:wigner}

Since the coherent theory of the Talbot-Lau effect can be found in the
literature \cite{Talbot1836a,Patorski1989a,Dubetsky1997a,Brezger2003a} we
shall present no detailed derivations, but discuss the approximations involved
and state the results in terms of the Wigner function as far as needed for the
later inclusion of decoherence effects. Consider the usual interferometric
situation where a flux of particles enters at $z=0$ with a longitudinal
momentum $p_{z}$ that is much greater than its transverse components. Ideally,
the particle is in a momentum eigenstate, or in an incoherent mixture thereof,
before passing a number of collimation slits and gratings. The vector
$\mathbf{r}=\left(  x,y\right)  $ describes the distance of the particles from
the interferometer axis. In the usual paraxial approximation this separation,
as well as the structures in the grating and in the collimation planes, are
assumed to be small compared to the distances $L_{i}$ between the optical
elements, $|\mathbf{r}|\ll L_{i}$. In this case one may evaluate the
transmission to leading order in $|\mathbf{r}|/L_{i}$. This approximation
implies that the longitudinal and the transverse part of the state remain
separable throughout the interferometer. It follows that the discussion may be
confined to the transverse degrees of freedom as described by $\psi
(\mathbf{r})$ if the evolution is completely coherent --- or, in the general
case, by the density matrix $\rho(\mathbf{r},\mathbf{r}^{\prime})$.

Given the wave function $\psi_{0}(\mathbf{r})$ on the $z=0$ plane the free
unitary evolution up to the plane $z=L$ yields, to leading order, the
transverse state,%
\begin{widetext}
\begin{equation}
\psi_{L}(\mathbf{r})=\frac{p_{z}}{2\pi\hbar\mathrm{i}L}\,\mathrm{e}%
^{\mathrm{i}p_{z}L/\hbar}\int\mathrm{d}\mathbf{r}_{0}\,\exp\!\left(
\mathrm{i}\frac{p_{z}}{\hbar}\frac{|\mathbf{r}-\mathbf{r}_{0}|^{2}}%
{2L}\right)  \psi_{0}(\mathbf{r}_{0})\,+{\mathcal{O}}\left(  \frac
{\mathbf{r}^{2}}{L^{2}}\right)  \;, \label{eq:parapr}%
\end{equation}
as follows from an asymptotic expansion of the free Green function,
e.g.\ \cite{Sommerfeld1950a}. An important feature of this paraxial
approximation is the fact that it reflects the composition property of the
exact propagation without any loss of accuracy. That is, propagating the wave
function first by a distance $L_{1}$ and subsequently by the distance $L_{2}$
yields exactly the same result as a single propagation by $L_{1}+L_{2}$,
\begin{align}
\psi_{3}(\mathbf{r}_{3})=  &  -\frac{p_{z}^{2}}{\left(  2\pi\hbar\right)
^{2}L_{1}L_{2}}\,\mathrm{e}^{\mathrm{i}\left(  p_{z}L_{1}+p_{z}L_{2}\right)
/\hbar}
\int\mathrm{d}\mathbf{r}_{1}\mathrm{d}\mathbf{r}_{2}\,\exp\!\left(
\mathrm{i}\frac{p_{z}}{\hbar}\frac{|\mathbf{r}_{3}-\mathbf{r}_{2}|^{2}}%
{2L_{2}}\right)  \exp\!\left(  \mathrm{i}\frac{p_{z}}{\hbar}\frac
{|\mathbf{r}_{2}-\mathbf{r}_{1}|^{2}}{2L_{1}}\right)  \psi_{1}(\mathbf{r}%
_{1})\nonumber\\
=  &  \frac{p_{z}}{2\pi\hbar\mathrm{i}(L_{1}+L_{2})}\,\mathrm{e}%
^{\mathrm{i}p_{z}(L_{1}+L_{2})/\hbar}\int\mathrm{d}\mathbf{r}_{1}%
\,\exp\!\left(  \mathrm{i}\frac{p_{z}}{\hbar}\frac{|\mathbf{r}_{1}%
-\mathbf{r}_{3}|^{2}}{2(L_{1}+L_{2})}\right)  \psi_{1}(\mathbf{r}_{1})\;,
\label{psi3}%
\end{align}
\end{widetext}
which follows from Gaussian integration. Hence, within the paraxial
approximation no loss of accuracy is introduced by dividing the propagation
into a sequence of intervals and integrating over the interjacent planes. This
freedom will be used below as a crucial ingredient when we describe the
effects of decoherence.

Note that the composition property (\ref{psi3}) does not require a large
separation between the planes. Even for infinitesimally close planes one
obtains the correct expression (\ref{eq:parapr}). This can be seen
immediately in (\ref{psi3}) by noting a particular representation of the
two-dimensional $\delta$ function \cite[eq.~(A.33)]{Hornberger2002b},
\begin{equation}
\frac{p_{z}}{2\pi\hbar\mathrm{i}L}\exp\!\left(  \mathrm{i}\frac{p_{z}}{\hbar
}\frac{|\mathbf{r}-\mathbf{r}_{0}|^{2}}{2L}\right)
\begin{CD} @>\hbar L/p_z\to0>>\delta_2(\mathbf{r}-\mathbf{r}_0) \end{CD}\;.
\end{equation}
\qquad

The first equality in (\ref{psi3}) also shows how the existence of an ideal
grating at $z=L_{1}$ would affect the propagated state. In the case of a
binary grating the $\mathbf{r}_{2}$-integration would be simply restricted to
the transparent parts of the grating plane. In general, an ideal grating
causes an amplitude and phase modulation
\begin{equation}
\psi_{1}^{\prime}(\mathbf{r}_{2})=t(\mathbf{r}_{2})\psi_{1}(\mathbf{r}%
_{2})\quad\text{with}\quad|t(\mathbf{r}_{2})|<1\;, \label{eq:gratingtrafo1}%
\end{equation}
which is accounted for by the additional appearance of a grating function
$t(\mathbf{r}_{2})$ under the integral.

The passage of a particle stream through a general interferometer may now be
described as a sequence of transmissions through gratings or collimation slits
as described by (\ref{eq:gratingtrafo1}) each followed by a free evolution
(\ref{eq:parapr}). This holds also for general, mixed states since any density
operator can be represented as a convex sum of projectors to pure states.

\subsection{Wigner function}

We proceed to formulate the propagation in the Wigner representation, which
has several advantages. First, the Wigner function permits a direct comparison
of the quantum evolution to the classical dynamics in terms of phase space
distributions. Second, and more importantly, it is the most convenient
starting point to include effects of decoherence in Sect.~\ref{sec:deco}.
Finally, the free evolution (\ref{eq:parapr}) has a particularly simple form
in the Wigner representation.

The Wigner function is the Fourier transformation of the position density
matrix $\rho(\mathbf{r},\mathbf{r}^{\prime})$ with respect to the two-point
separation $\boldsymbol{\Delta}=\mathbf{r}-\mathbf{r}^{\prime}$
\cite{Wigner1932a},
\begin{equation}
w(\mathbf{r},\mathbf{p})=\frac{1}{(2\pi\hbar)^{2}}\int\mathrm{d}%
\boldsymbol{\Delta}\,\mathrm{e}^{\mathrm{i}\mathbf{p}\boldsymbol{\Delta/}%
\hbar}\rho\left(  \mathbf{r}-\frac{\boldsymbol{\Delta}}{2},\mathbf{r}%
+\frac{\boldsymbol{\Delta}}{2}\right)  \;. \label{eq:wignerdef}%
\end{equation}
It may be viewed as a quantum analogue to the classical phase space
distribution $f(\mathbf{r},\mathbf{p})$, with $\mathbf{p}$ the transverse
momentum vector.

In order to obtain the free unitary evolution of the Wigner function we note
that the density matrix in position representation has the general form
\begin{equation}
\rho(\mathbf{r},\mathbf{r}^{\prime})=\int\mathrm{d}\mu\,g(\mu)\psi_{\mu
}(\mathbf{r})\psi_{\mu}^{\ast}(\mathbf{r}^{\prime})
\end{equation}
with $\int\mathrm{d}\mu\,g(\mu)=1$. According to (\ref{eq:parapr}) a free
unitary evolution by the distance $L$ yields
\begin{align}
&\rho(\mathbf{r},\mathbf{r}^{\prime})=\frac{p_{z}^{2}}{\left(  2\pi
\hbar\right)  ^{2}L^{2}}\nonumber
\\
&\times
\int\mathrm{d}\mathbf{r}_{0}\mathrm{d}\mathbf{r}%
_{0}^{\prime}\,\exp\!\left(  \mathrm{i}\frac{p_{z}}{\hbar}\frac{|\mathbf{r}%
-\mathbf{r}_{0}|^{2}-|\mathbf{r}^{\prime}-\mathbf{r}_{0}^{\prime}|^{2}}%
{2L}\right)  \rho_{0}(\mathbf{r}_{0},\mathbf{r}_{0}^{\prime}%
)\;.\label{eq:rhotrafo}%
\end{align}
From (\ref{eq:wignerdef}) and (\ref{eq:rhotrafo}) it follows that a free
unitary evolution by the distance $L$ changes the Wigner function according
to
\begin{equation}
w_{L}(\mathbf{r},\mathbf{p})=w_{0}\left(  \mathbf{r}-\frac{L}{p_{z}}%
\mathbf{p},\mathbf{p}\right)  \;.\label{eq:freetrafo1}%
\end{equation}
This transformation is particularly simple and, as one expects, is identical
to the free movement of a classical probability density in the phase space of
the transverse degree of freedom. The decisive difference between the
classical and the quantum phase space dynamics is found in the transformation
for passing through a grating. Equation (\ref{eq:gratingtrafo1}) implies that
by going through a grating the Wigner function undergoes a convolution
\begin{equation}
w^{\prime}(\mathbf{r},\mathbf{p})=\int\mathrm{d}\mathbf{q}\,T(\mathbf{r}%
,\mathbf{q})w(\mathbf{r},\mathbf{p}-\mathbf{q})\label{eq:gratingtrafo2x}%
\end{equation}
which in general builds up quantum coherences that show up as oscillations in
the momentum direction. Here we define the convolution kernel analogously to
(\ref{eq:wignerdef}) as
\begin{equation}
T(\mathbf{r},\mathbf{p})=\frac{1}{(2\pi\hbar)^{2}}\int\mathrm{d}%
\boldsymbol{\Delta}\,\mathrm{e}^{\mathrm{i}\mathbf{p}\boldsymbol{\Delta/}%
\hbar}t\left(  \mathbf{r}-\frac{\boldsymbol{\Delta}}{2}\right)  t^{\ast
}\left(  \mathbf{r}+\frac{\boldsymbol{\Delta}}{2}\right)  \;.\label{eq:Tdef}%
\end{equation}
Note that by stating (\ref{eq:gratingtrafo2x}) we do not keep the
normalization of the Wigner function. Indeed, a finite fraction of the
particles may hit the grating and may be removed from the flux. Therefore it
is convenient to work with an unnormalized state and take care of the
normalization only in the end.

With the transformations (\ref{eq:freetrafo1}) and (\ref{eq:gratingtrafo2x})
we can proceed to describe the Talbot-Lau effect in a general framework.

\subsection{The Talbot-Lau setup}

\label{sec:TLsetup}

In the Talbot-Lau setup a monochromatic beam passes three vertical gratings
that are separated by the distances $L_{1}$ and $L_{2}$. Since the particle
stream is effectively uncollimated in front of the first grating its Wigner
function for the transverse degrees of freedom is uniform. If we start with
the (improper) normalization $w_{0}(\mathbf{r},\mathbf{p})=1$ then
(\ref{eq:Tdef}) yields the Wigner function after the first grating
\begin{equation}
w_{1}(\mathbf{r},\mathbf{p})=|t_{1}(\mathbf{r})|^{2}\;.
\end{equation}
The free unitary evolution by a distance $L_{1}$, followed by a passage
through a grating (with convolution kernel $T$) and another free evolution by
a distance $L_{2},$ leads to the general expression%

\begin{multline}
w(\mathbf{r},\mathbf{p})=\,\int\mathrm{d}\mathbf{q}\,\left\vert t_{1}\left(
\mathbf{r}-\frac{\mathbf{p}}{p_{z}}(L_{1}+L_{2})+\frac{\mathbf{q}}{p_{z}}%
L_{1}\right)  \right\vert ^{2}\\\times T\left(  \mathbf{r}-\frac{\mathbf{p}}{p_{z}%
}L_{2}\,,\mathbf{q}\right)  \;. \label{eq:w3general}%
\end{multline}
The particle density at position $z=L_{1}+L_{2}$ is obtained by integrating
the momentum variable. It can be written as
\begin{equation}
w(\mathbf{r})\equiv\,\int w(\mathbf{r},\mathbf{p})\,\mathrm{d}\mathbf{p}%
=\int\mathrm{d}\mathbf{r}_{1}\,|t_{1}(\mathbf{r}_{1})|^{2}\,h(\mathbf{r}%
;\mathbf{r}_{1}) \label{eq:wxgeneral}%
\end{equation}
with
\begin{multline}
h(\mathbf{r};\mathbf{r}_{1})=\left(  \frac{p_{z}}{\hbar L_{1}}\right)
^{2}\\\times \int\mathrm{d}\mathbf{p}\,T\left(  \mathbf{r}-\frac{\mathbf{p}}{p_{z}%
}L_{2}\,,\frac{L_{1}+L_{2}}{L_{1}}\mathbf{p}-\frac{\mathbf{r}-\mathbf{r}_{1}%
}{L_{1}}p_{z}\right)  \;. \label{eq:hdef}%
\end{multline}

\begin{figure}
[ptb]
\begin{center}
\includegraphics[
width=\columnwidth
]%
{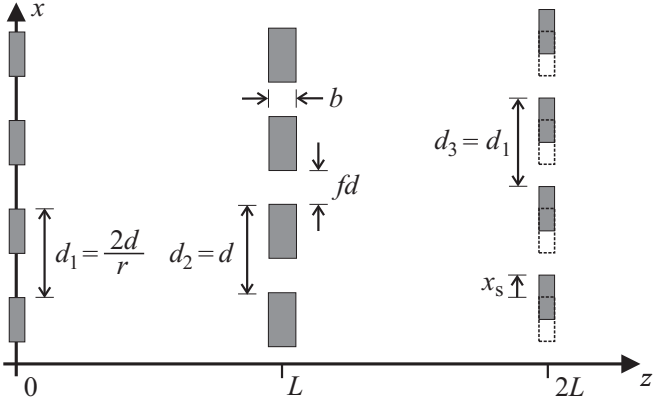}%
\caption{The symmetric Talbot-Lau setup consists of three parallel gratings
separated by equal distances $L$. Near-field interference of an uncollimated
beam from the left may lead to a density pattern at the position of the third
grating that can be observed by modulation with the lateral grating position
$x_{\mathrm{s}}.$}%
\label{drawing}%
\end{center}
\end{figure}

As mentioned above the Talbot-Lau effect operates in the near-field regime,
where the fact that the gratings have a finite lateral extension does not play
a role; it only affects the overall count rate. It is therefore permissible to
describe the gratings by idealized functions that are periodic on an infinite
plane. Moreover, since the setup is invariant with respect to changes in the
vertical position, it is sufficient to consider the Wigner function and the
grating transmissions only as a function of the \textquotedblleft
horizontal\textquotedblright\ coordinates $(r_{x},p_{x})\equiv(x,p)$, see
Fig.\ \ref{drawing}.%

We take the first grating to have a period $d_{1}$, and its transmission
function to be given by $t_{1}(x)=\sum a_{m}\exp(2\pi\mathrm{i}mx/d_{1})$.
Likewise, the second grating has the Fourier coefficients $b_{m}$ and the
period $d_{2}=d$. Therefore,
\begin{equation}
|t_{1}(x)|^{2}=\sum_{\ell\in\mathbb{Z}}A_{\ell}\exp\!\left(  2\pi
\mathrm{i}\ell\frac{x}{d_{1}}\right)  \quad\text{with } A_{\ell}%
=\sum_{j\in\mathbb{Z}}a_{j}a_{j-\ell}^{\ast} \label{eq:t1sqfourier}%
\end{equation}
{and}
\begin{equation}
T(x,p)=\sum_{\ell,j\in\mathbb{Z}}b_{j}b_{j-\ell}^{\ast}\exp\!\left(
2\pi\mathrm{i}\ell\frac{x}{d}\right)  \delta\left(  p-\hbar\pi\frac{2j-\ell
}{d}\right)  . \label{eq:Tfourier}%
\end{equation}

In order to simplify the discussion and to avoid an overly complicated
notation we focus on the \emph{symmetric} Talbot-Lau setup, which is the most
important one in practice; for the asymmetric setup see \cite{Brezger2003a}
and note \footnote{To observe the \emph{asymmetric }Talbot-Lau effect, at
$L_{2}=\kappa L_{1}$ with $\kappa>0$, a period of $d_{1}=\left(
\kappa+1\right)  /\kappa\times d/r$ is needed in the first grating.
Analogously to the symmetric case (\ref{eq:qdensity}) a density pattern
emerges at $z=L_{1}+L_{2}$ which has now a period of $\kappa d_{1}$. The case
$\kappa<1$ is called the \emph{fractional }Talbot-Lau effect, and it is easily
incorporated into the present framework.}. In this case the gratings are set
at an equal distance $L_{1}=L_{2}\equiv L$ and the periods $d_{1}$ and $d$ of
the first and second grating are related by $d_{1}=2d/r$, with $r\in
\mathbb{N}$. (The case of equal grating periods, $r=2$, is most common
\cite{Brezger2002a,Hornberger2003a,Hackermuller2003a,Hackermuller2004a}.) The
state (\ref{eq:w3general}) now reads%
\begin{align}
w(x,p)=&\frac{1}{\hbar}\sum_{\ell,j,m\in\mathbb{Z}}   A_{\ell}b_{j}%
b_{j-m}^{\ast}\nonumber
\\\nonumber 
&\times \exp\!\left(  2\pi\mathrm{i}\left(  \frac{r}{2}\ell+m\right)
\frac{x}{d}-2\pi\mathrm{i}(r\ell+m)\frac{L}{d}\frac{p}{p_{z}}\right)
\label{eq:w3}\\
&  \times\exp\!\left(  \mathrm{i}\pi\ell(2j-m)\frac{r}{2}\frac{L}{{L_{\lambda
}}}\right)  \;.
\end{align}
Here we introduced the \emph{Talbot length}%

\begin{equation}
{L_{\lambda}}=\frac{d^{2}p_{z}}{2\pi\hbar}=\frac{d^{2}}{\lambda}%
\end{equation}
in terms of the grating period $d$ and the de Broglie wavelength
$\lambda=h/p_{z}.$ The Talbot length is the proper scale to distinguish
near-field (Fresnel) diffraction from far-field (Fraunhofer) diffraction for a
given wavelength. It gives the distance behind a grating where the diffraction
peaks of a collimated, passing beam have a lateral separation equal to the
grating period $d$.

To get the particle density in the beam from (\ref{eq:w3}) we integrate over
the momentum $p,$ which picks up the $m=-r\ell$ components \footnote{In order
to specify the proportionality factor in (\ref{eq:qdensity}) a normalizable
momentum distribution would be needed in the initial Wigner function.},
\begin{equation}
w(x)\equiv\int\mathrm{d}p\,w(x,p)\,\propto\,\sum_{\ell\in\mathbb{Z}}A_{\ell
}^{\ast}B_{\ell r}^{(\lambda)}\exp\!\left(  2\pi\mathrm{i}\ell\frac{x}{d_{1}%
}\right)  \label{eq:qdensity}%
\end{equation}
with Fourier components \cite{Brezger2003a}
\begin{equation}
B_{m}^{(\lambda)}=\sum_{j\in\mathbb{Z}}b_{j}b_{j-m}^{\ast}\exp\!\left(
\mathrm{i}\pi\frac{m^{2}-2jm}{2}\frac{L}{{L_{\lambda}}}\right)
~.\label{eq:Bdefx}%
\end{equation}
Equation (\ref{eq:qdensity}) predicts a density pattern which has the same
period $d_{1}$ as the first grating. Often the spatial resolution of detectors
is too poor to detect these density oscillations directly in an experiment.
However, an indirect observation is possible with the help of a third grating
with period $d_{1}$. If put at the position of the density pattern it
modulates the total transmission as a function of its lateral position
$x_{\mathrm{s}}$. The integrated transmission, which is much easier to detect,
is then given by
\begin{align}
S(x_{\mathrm{s}})&=\int\mathrm{d}p\,\mathrm{d}q\,\mathrm{d}%
x\,w(x,p-q)T(x-x_{\mathrm{s}},q)\nonumber\\
&=\int\mathrm{d}x\,w(x)|t_{3}(x-x_{\mathrm{s}%
})|^{2}\;.\label{eq:qsignal1}%
\end{align}
If we choose the first and third grating to be identical, $t_{3}(x)=t_{1}(x)$,
the expected periodic signal is given by the expression
\begin{equation}
S(x_{\mathrm{s}})\,\propto\,\sum_{\ell\in\mathbb{Z}}(A_{\ell}^{\ast}%
)^{2}B_{\ell r}^{(\lambda)}\exp\!\left(  2\pi\mathrm{i}\ell\frac
{x_{\mathrm{s}}}{d_{1}}\right)  \;.\label{eq:qsignal2}%
\end{equation}
For \emph{symmetric} gratings this modulation signal has a visibility
\footnote{Equations (\ref{eq:qvisibility}) and (\ref{eq:cvisibility}) assume
that the transmission signal is extremal at $x_{\mathrm{s}}=0$ and
$x_{\mathrm{s}}=d_{1}/2$ which is the case for realistic transmission
functions with $t(x)=t(-x)$.}
\begin{equation}
\mathcal{V}_{\mathrm{qm}}=\frac{S_{\mathrm{max}}-S_{\mathrm{min}}%
}{S_{\mathrm{max}}+S_{\mathrm{min}}}=\frac{\left\vert \sum\limits_{n=1}%
^{\infty}A_{2n-1}^{2}B_{2nr-r}^{(\lambda)}\right\vert }{\frac{1}{2}A_{0}%
^{2}B_{0}^{(\lambda)}+\sum\limits_{n=1}^{\infty}A_{2n}^{2}B_{2nr}^{(\lambda)}%
}\label{eq:qvisibility}%
\end{equation}
which serves as the prime characterization of the interference pattern.

It is clear from the definition of the coefficients $B_{m}^{(\lambda)}$ that
the interference pattern (\ref{eq:qdensity}) and the visibility
(\ref{eq:qvisibility}) depend strongly on both the wavelength $\lambda$ and on
the separation $L$ between the gratings. As evident from (\ref{eq:Bdefx}) it
is indeed the product of the two quantities which determines the pattern,
since $L/{L_{\lambda}}=L\lambda/d^{2}$.

However, the detection of a periodic signal alone does not prove necessarily
that quantum interference occurred in the experiment because a certain density
pattern may also be expected from a generalized Moir\'{e} effect. To establish
the observation of quantum interference one must show that the observed
visibility differs significantly from the classical expectation. It is
therefore important to have a reliable quantitative prediction for the
classical expectation as well.

\subsection{The classical expectation}

\label{sec:class}

With the results for the Wigner function at hand it is straightforward to
repeat the calculation using classical phase space dynamics. The classical
phase space density $f(\mathbf{r},\mathbf{p})$ transforms under free evolution
like the Wigner function (\ref{eq:freetrafo1}) according to%

\begin{equation}
f_{L}(\mathbf{r},\mathbf{p})=f_{0}\left(  \mathbf{r}-\frac{L}{p_{z}}%
\mathbf{p},\mathbf{p}\right)  \;.\label{eq:freetrafo2}%
\end{equation}
In contrast to (\ref{eq:Tdef}), the convolution kernel for passing through an
ideal amplitude grating is now given by
\begin{equation}
T_{\mathrm{cl}}(\mathbf{r},\mathbf{p})=\frac{1}{(2\pi\hbar)^{2}}\int
\mathrm{d}\boldsymbol{\Delta}\,\mathrm{e}^{\mathrm{i}\mathbf{p}%
\boldsymbol{\Delta/}\hbar}|t(\mathbf{r})|^{2}=|t(\mathbf{r})|^{2}%
\delta(\mathbf{p})\;,\label{eq:Tcldef}%
\end{equation}
which leads to
\begin{equation}
f^{\prime}(\mathbf{r},\mathbf{p})=|t(\mathbf{r})|^{2}f(\mathbf{r}%
,\mathbf{p})\;.\label{eq:gratingtrafo3}%
\end{equation}
At a distance $L_{2}$ after the second grating this yields a phase space
distribution
\begin{equation}
f(\mathbf{r},\mathbf{p})=\,\left\vert t_{1}\left(  \mathbf{r}-\frac
{\mathbf{p}}{p_{z}}(L_{1}+L_{2})\right)  \right\vert ^{2}\,\left\vert
t_{2}\left(  \mathbf{r}-\frac{\mathbf{p}}{p_{z}}L_{2}\right)  \right\vert ^{2}%
\end{equation}
which can also be obtained from the quantum result by replacing $T$ by
$T_{\mathrm{cl}}$ in (\ref{eq:w3general}). It follows that the classical
density pattern in front of the third grating is given by
\begin{equation}
f(x)\equiv\int\mathrm{d}p\,f(x,p)\,\propto\,\sum_{\ell\in\mathbb{Z}}A_{\ell
}^{\ast}B_{\ell r}^{(0)}\exp\!\left(  2\pi\mathrm{i}\ell\frac{x}{d_{1}%
}\right)  \label{eq:cdensity}%
\end{equation}
with
\begin{equation}
B_{m}^{(0)}=\sum_{j\in\mathbb{Z}}b_{j}b_{j-m}^{\ast}\;.\label{eq:B0def}%
\end{equation}
The comparison with (\ref{eq:qdensity}) shows that the quantum and the
classical results have the same form, but differ in the Fourier components
$B_{m}$. Of course, the classical Fourier components do not depend on the de
Broglie wavelength. Nonetheless, the $B_{m}^{(0)}$ may be viewed as the
short-wave limit $L/{L_{\lambda}}\rightarrow0$ of the quantum Fourier
coefficients $B_{m}^{(\lambda)}$, which is already indicated by the notation.

It follows immediately that the classical prediction for the signal is
obtained from (\ref{eq:qsignal2}) by replacing the wavelength dependent
Fourier components $B_{m}^{(\lambda)}$ by the $B_{m}^{(0)}$,%
\begin{equation}
S_{\text{cl}}(x_{\mathrm{s}})\,\propto\,\sum_{\ell\in\mathbb{Z}}(A_{\ell
}^{\ast})^{2}B_{\ell r}^{(0)}\exp\!\left(  2\pi\mathrm{i}\ell\frac
{x_{\mathrm{s}}}{d_{1}}\right)  \;. \label{csignal}%
\end{equation}
Likewise one can show that the visibility of the classical signal is given,
for symmetric gratings, by
\begin{equation}
\mathcal{V}_{\mathrm{cl}}=\frac{\left\vert \sum\limits_{n=1}^{\infty}%
A_{2n-1}^{2}B_{2nr-r}^{(0)}\right\vert }{\frac{1}{2}A_{0}^{2}B_{0}^{(0)}%
+\sum\limits_{n=1}^{\infty}A_{2n}^{2}B_{2nr}^{(0)}}\;. \label{eq:cvisibility}%
\end{equation}

Let us focus on the important case of equal grating periods for a moment,
i.e., $r=2$. In this case only the even Fourier components $B_{2m}$ are
needed. If the separation $L$ between the gratings is set to an integer
multiple of the Talbot length, then it is easy to convince oneself that for
any ideal grating $t_{2}(x)$
\begin{equation}
B_{2m}^{(\lambda)}=B_{2m}^{(0)}\qquad\text{if\ }\;L/{L_{\lambda}}\in\mathbb{N}
\label{eq:theyareequal}%
\end{equation}
that is, the quantum and the classical evolution yield \emph{identical}
predictions for the density pattern and the observed visibility. This shows
clearly that the observation of the integer Talbot-Lau effect alone does not
prove the wave nature of the beam particles.

\begin{figure}
[ptb]
\begin{center}
\includegraphics[
width=\columnwidth
]%
{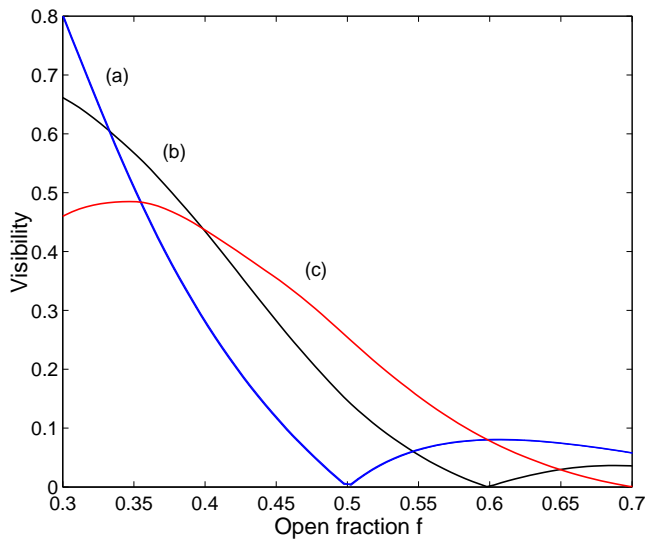}%
\caption{Talbot-Lau visibilities for ideal binary gratings as a function of
the open fraction. (a) $L=L_{\lambda}$ (b) $L=0.9L_{\lambda}$ (c)
$L=0.8L_{\lambda}$. The corresponding classical visibilities are all identical
with curve (a).}%
\label{fig:idealvisib}%
\end{center}
\end{figure}

However, unlike their classical counterparts, the quantum Fourier components
display a strong wavelength dependence. Therefore, distinctively different
results are obtained in the classical and the quantum calculation for
separations which differ from the integer Talbot criterion, $L/{L_{\lambda}%
}\notin\mathbb{N}$, or equivalently for detuned particle wavelengths,
$\lambda\neq nd^{2}/L,n\in\mathbb{N}$. This can be seen in
Fig.~\ref{fig:idealvisib} where we show the quantum and classical visibilities
for identical, ideal binary gratings as a function of their open fraction $f$
(the ratio of slit width to grating period).%

As predicted by (\ref{eq:theyareequal}) the quantum and classical results are
identical for $L/{L_{\lambda}}=1$ and given by Fig.~\ref{fig:idealvisib}(a).
For $L/{L_{\lambda}}=0.9$ (Fig.~\ref{fig:idealvisib}(b)) and $L/{L_{\lambda}%
}=0.8$ (Fig.~\ref{fig:idealvisib}(c)) on the other hand, the Talbot-Lau
visibilities differ markedly while the classical predictions remain on
Fig.~\ref{fig:idealvisib}(a). The distinction between the classical and the
quantum predictions is most pronounced for an opening fraction of $0.5$, where
the classical contrast vanishes. The quantum calculation yields significant
visibilities for these gratings, of $14.7\%$ at $L/{L_{\lambda}}=0.9$ and
$25.4\%$ at $L/{L_{\lambda}}=0.8$, respectively.

\subsection{Finite longitudinal coherence}

\label{sec:lcoh}In the above calculations the particle beam that enters the
interferometer was assumed to have a fixed velocity in the $z$-direction and
to be completely uncollimated in the transverse direction. Of course this is
an idealization that is in many respects as unrealistic as the familiar
assumption of a perfectly coherent plane wave. Realistic particle beams are
characterized by a finite longitudinal coherence and show some correlations in
the transverse direction.

The particle beams used in matter wave interferometry are usually generated by
an effusive or supersonic expansion into a vacuum chamber \cite{Scoles1988a}.
By means of additional skimmers or collimators the beam is restricted to a
well-defined \textquotedblleft longitudinal\textquotedblright\ direction. The
beam is stationary, and as a consequence the longitudinal momenta show no
off-diagonal elements in their density matrix \cite{Rubenstein1999b}. They are
completely characterized by the longitudinal momentum distribution $g(p_{z})$
\cite{Englert1994a}.

The transverse momenta are much smaller than the longitudinal ones and can be
taken to be uncorrelated with the longitudinal velocity. The transverse
coherence is determined by the source aperture \cite{Born1980a}, and it could
be calculated with the van~Cittert-Zernike theorem \cite{Born1980a} if the
aperture size were small compared to the length scale in question. However, in
the Talbot-Lau setup the source aperture is much larger than the spacing
between the grating slits. As a consequence, diffraction at the first grating
cannot be observed and it is permissible to approximate the transverse degrees
of freedom as completely uncollimated. In front of the first grating the bulk
of the beam is therefore appropriately characterized by the Wigner function,
\begin{equation}
W_{\mathrm{beam}}(\mathbf{R},\mathbf{P})=g(\mathbf{P}\cdot\mathbf{e}%
_{z}),\label{eq:w3d}%
\end{equation}
for $\mathbf{R}=\mathbf{r}+z\mathbf{e}_{z}$ with $z<0.$ This description does
not account for the edges of the beam and the cut-off at larger transverse
momenta, which is why it cannot be properly normalized. Fortunately, it is not
necessary to include the full beam profile in the treatment of the Talbot-Lau
effect. As shown below only the interference of paths through a few
neighboring slits is relevant for the effect, so that the transverse variation
of the total current can be neglected.

Formally, the beam (\ref{eq:w3d}) can be written as a convex sum of states
$w(\mathbf{r},\mathbf{p})=1$ that are uniform in the transverse coordinates
$\mathbf{r}$ and $\mathbf{p}$, and that have a fixed longitudinal momentum
$p_{z}$.
\begin{equation}
W_{\mathrm{beam}}(\mathbf{r}+z\mathbf{e}_{z},\mathbf{p}+p_{z}\mathbf{e}%
_{z})=\int_{0}^{\infty}\delta(p_{z}-p_{z}^{\prime})\,\underbrace
{w(\mathbf{r},\mathbf{p})}_{=1}\,g(p_{z}^{\prime})\,\mathrm{d}p_{z}^{\prime}.
\label{eq:w3d2}%
\end{equation}
Those are the states $w(\mathbf{r},\mathbf{p})=1$ that we started out with in
Sect.~\ref{sec:TLsetup}. Since a sequence of grating transmissions
(\ref{eq:gratingtrafo2x}) and free evolutions (\ref{eq:freetrafo1}) does not
affect the dependence on $p_{z}$ the general stationary state at $z^{\prime
}>0$ is given by%
\begin{equation}
W_{\mathrm{beam}}(\mathbf{r}+z^{\prime}\mathbf{e}_{z},\mathbf{p}%
+p_{z}\mathbf{e}_{z})=w(z^{\prime};\mathbf{r},\mathbf{p})g\left(
p_{z}\right)
\end{equation}
and the transverse position density reads%
\begin{align}
w\left(  \mathbf{r}\right)   &  =\int\mathrm{d}\mathbf{p}\mathrm{d}%
p_{z}W_{\mathrm{beam}}(\mathbf{r}+z^{\prime}\mathbf{e}_{z},\mathbf{p}%
+p_{z}\mathbf{e}_{z})\nonumber\\
&  =\int\mathrm{d}\mathbf{p}\mathrm{d}p_{z}~w(z^{\prime};\mathbf{r}%
,\mathbf{p})g\left(  p_{z}\right)  .
\end{align}
It follows that the finite longitudinal coherence in the beam is completely
accounted for by averaging the results for a fixed velocity derived in
Sects.~\ref{sec:TLsetup} and \ref{sec:class} over the longitudinal velocity
distribution. In particular, the modulation signals (\ref{eq:qsignal2}) and
(\ref{csignal}) are given by
\begin{equation}
\langle S(x_{\mathrm{s}})\rangle=\int_{0}^{\infty}\mathrm{d}p_{z}%
~g(p_{z})S(x_{\mathrm{s}}). \label{eq:meanS}%
\end{equation}
If the detection signal is proportional to the flux the average involves the
longitudinal velocity component, ~%
\begin{equation}
\langle S(x_{\mathrm{s}})\rangle=\frac{\int_{0}^{\infty}\mathrm{d}%
p_{z}~g(p_{z})p_{z}S(x_{\mathrm{s}})}{\int_{0}^{\infty}\mathrm{d}p_{z}%
~p_{z}g(p_{z})}.
\end{equation}
Note that in either case the visibilities are not obtained by a simple average
of (\ref{eq:qvisibility}) and (\ref{eq:cvisibility}), but have to be
calculated from the averaged signal.

\section{The influence of realistic gratings}

\label{sec:gratings}

So far it has been assumed that the gratings are ideal in the sense that their
thickness could be neglected. However, real gratings have a {finite} thickness
and the time of interaction between the particle and the grating depends on
the velocity $v_{z}=p_{z}/m_{\mathrm{p}}$ of the beam particles. This
introduces a velocity dependence of the grating function both in the classical
and in the quantum treatments. Generally speaking, the Talbot-Lau effect is
affected more strongly by the grating forces than far-field diffraction
\cite{Bruehl2002a}, since the near-field interference is characterized by
smaller phase shifts. This was seen in recent experiments with beams of large
molecules
\cite{Brezger2002a,Hornberger2003a,Hackermuller2003a,Hackermuller2004a}.

\subsection{The grating interaction}

In order to account for the effect of a finite grating thickness $b$ we
consider an additional interaction potential $V(x)$ that acts while the
particle is traversing the grating. In order to avoid a more detailed
$z$-dependence in the potential we average over the the surface roughness and
assume that the grating walls are parallel to the optical axis and that edge
effects can be neglected. In the case of tilted walls one can introduce an
effective slit width, as discussed in \cite{Stoll2003a}. Moreover, it is known
\cite{Grisenti1999a,Bruhl2002a} that grating interaction effects are usually
well described by the \emph{eikonal approximation}. There, the additional
quantum phase due to the interaction potential $V(x)$ is obtained by
integrating the action along a straight path. Accordingly, if we take the
binary function $t(x)$ to describe the material grating the complete grating
function is given by
\begin{equation}
\tilde{t}(x)=t(x)\exp\!\left(  -\mathrm{i}\frac{m_{\mathrm{p}}b}{p_{z}}%
\frac{V(x)}{\hbar}\right)  . \label{eq:qeikonal}%
\end{equation}
Here and below the tilde is used to indicate quantities which have an
additional velocity dependence due to the grating interaction. Accordingly,
for non-ideal gratings the convolution kernel (\ref{eq:Tdef}) is replaced by
\begin{equation}
\widetilde{T}(x,p)=\int\mathrm{d}q\,T(x,p-q)T_{V}(x,q) \label{eq:Ttildedef}%
\end{equation}
with
\begin{multline}
T_{V}(x,q)=\frac{1}{2\pi\hbar}\int\mathrm{d}\Delta\,\mathrm{e}^{\mathrm{i}%
q\Delta/\hbar}\\\times
\exp\!\left(  -\mathrm{i}\frac{m_{\mathrm{p}}b}{p_{z}\hbar
}\left[  V\left(  x-\frac{\Delta}{2}\right)  -V\left(  x+\frac{\Delta}%
{2}\right)  \right]  \right)  \;. \label{eq:Vkernel}%
\end{multline}
It follows that the quantum expressions (\ref{eq:qdensity}) for the density
pattern and (\ref{eq:qvisibility}) for the signal visibility still hold after
the replacement%

\begin{equation}
B_{m}^{(\lambda)}\rightarrow\widetilde{B}_{m}^{(\lambda)}=\sum_{j\in
\mathbb{Z}}\tilde{b}_{j}\tilde{b}_{j-m}^{\ast}\exp\!\left(  \mathrm{i}\pi
\frac{m^{2}-2jm}{2}\frac{L}{{L_{\lambda}}}\right)  \label{eq:Btilde}%
\end{equation}
with the modified Fourier components $\tilde{b}_{m}=\sum b_{j}\tilde{c}_{m-j}$
and
\begin{equation}
\tilde{c}_{m}=\frac{1}{d}\int_{-\frac{d}{2}}^{\frac{d}{2}}\mathrm{e}%
^{-2\pi\mathrm{i}m{x}/{d}}\exp\!\left(  -\mathrm{i}\frac{m_{\mathrm{p}}%
b}{p_{z}}\frac{V(x)}{\hbar}\right)  \mathrm{d}x\;.\label{eq:ctilde}%
\end{equation}
As mentioned above the presence of $V(x)$ introduces a velocity dependence
also on the classical level. This can be seen by considering the local
approximation of (\ref{eq:Vkernel}) where terms of the order $\partial_{x}%
^{3}V(x)\Delta^{3}$ are neglected in the exponent,%

\begin{align}
T_{V}(x,q)&\simeq\,\frac{1}{2\pi\hbar}\int\mathrm{d}\Delta\,\exp\!\left(
\mathrm{i}\left[  q+\frac{m_{\mathrm{p}}b}{p_{z}}\frac{\mathrm{d}}%
{\mathrm{d}x}V(x)\right]  \frac{\Delta}{\hbar}\right)  
\nonumber\\&=\delta\left(
q+\frac{m_{\mathrm{p}}b}{p_{z}}\frac{\mathrm{d}}{\mathrm{d}x}V(x)\right)
.\label{eq:Tshortrange}%
\end{align}
According to (\ref{eq:Ttildedef}) this yields a classical convolution kernel
\begin{align}
\widetilde{T}_{\mathrm{cl}}(x,p)&=|t_{2}(x)|^{2}\delta\left(  p+\frac
{m_{\mathrm{p}}b}{p_{z}}\frac{\mathrm{d}}{\mathrm{d}x}V(x)\right)
\nonumber\\&\equiv|t_{2}(x)|^{2}\delta(p-Q(x))\label{eq:Tcltilde}%
\end{align}
which indicates that the eikonal approximation corresponds on the classical
level to the momentum kick $Q(x)=-\partial_{x}V(x)\times b/v_{z}$ obtained by
multiplying the constant classical force at a \emph{fixed} position $x$ with
the interaction time. Accordingly, the classical phase space distribution
changes as $f^{\prime}(x,p)=|t(x)|^{2}f(x,p-Q(x))$ when passing a grating.
Using this transformation and the periodicity of $Q$ one finds that the
classical expressions for the density pattern and for the signal visibility
assume the forms (\ref{eq:cdensity}) and (\ref{eq:cvisibility}) as in the ideal
case. One merely has to replace the Fourier components by
\begin{equation}
{B}_{n}^{(0)}\rightarrow\widetilde{B}_{n}^{(0)}=\sum_{m\in\mathbb{Z}}{B}%
_{m}^{(0)}\widetilde{C}_{n-m}^{n}\label{eq:Bnulltilde}%
\end{equation}
with
\begin{equation}
\widetilde{C}_{m}^{n}=\frac{1}{d}\int_{-\frac{d}{2}}^{\frac{d}{2}}%
\mathrm{e}^{-2\pi\mathrm{i}m{x}/{d}}\exp\left(  -\mathrm{i}\pi n\frac{L}%
{d}\frac{Q(x)}{p_{z}}\right)  \mathrm{d}x\;.\label{eq:cdef}%
\end{equation}
We note that the modifications of the Fourier components given by
(\ref{eq:Btilde}) and (\ref{eq:Bnulltilde}) describe the quantum and the
classical interactions on the same degree of approximation. Clearly, the fact
that the interaction with the grating is treated equally in the quantum and in
the classical descriptions is an important requirement for identifying quantum
interference in an experimental observation.

Before turning to realistic descriptions for the grating interaction we note
that it is in general not necessary to include the grating interaction at the
first and third gratings. This is clear from the fact that the Talbot-Lau setup
is sensitive to diffraction only at the second grating, while the others
merely serve to modulate the flux. Formally, it can be seen from the
expression for the observed signal, Equation (\ref{eq:qsignal1}) with
(\ref{eq:wxgeneral}). It depends only on the squared moduli $|t_{1}(x)|^{2}$
and $|t_{3}(x)|^{2}$ of the first and third grating function, which are not
affected by the phase shift in (\ref{eq:qeikonal}). Only for very strong
potentials, where (\ref{eq:qeikonal}) is no longer valid, may the interaction 
effectively reduce the slit width and thus become relevant to the first and
third grating.

\subsection{Material gratings}

A neutral particle will in general experience an attractive van der Waals
force if placed in the vicinity of a surface. A simple, but quite realistic
description for nonpolar quantum objects is given by the static \emph{London
dispersion force} which acts between a quantum object and a flat wall. If
$\Delta$ is the distance to the wall it gives rise to the potential
$U(\Delta)=-C_{3}/\Delta^{3}$, with $C_{3}>0$ \cite{Dzyaloshinskii1961a}.

For simple wall materials the interaction constant $C_{3}$ can be found in the
literature for many atoms and a number of small molecules \cite{Vidali1991a}.
In general it is obtained from the Lifshitz formula
\cite{Dzyaloshinskii1961a,McLachlan1964a,Zaremba1976a}%

\begin{equation}
C_{3}=\frac{\hbar}{4\pi}\int_{0}^{\infty}\alpha(\mathrm{i}\omega
)\frac{\epsilon(\mathrm{i}\omega)-1}{\epsilon(\mathrm{i}\omega)+1}%
\mathrm{d}\omega\, \label{eq:C3Lifshitz}%
\end{equation}
by using either experimental data (e.g. absorption spectra) or appropriate
models for the dynamic polarizability $\alpha$ of the particle \footnote{We
follow common practice and take the polarizabilities in units of volume rather
than SI.\label{alphanote}} and for the bulk dielectric function $\epsilon$ of
the grating material, respectively. Often Drude-type models for $\alpha$ and
$\epsilon$ are considered sufficient. In \cite{Grisenti1999a,Bruhl2002a} the
interaction with material gratings was studied in an interference experiment
and found to be in good agreement with the assumption of a London dispersion force.

\begin{figure}
[ptb]
\begin{center}
\includegraphics[
width=\columnwidth
]%
{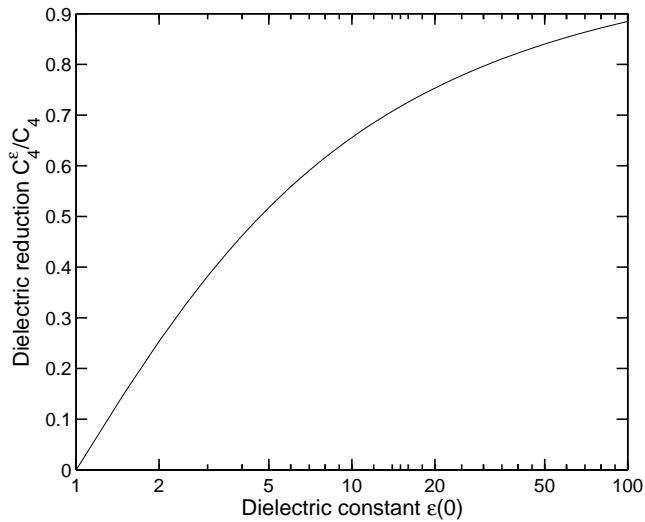}%
\caption{Reduction factor $C_{4}^{\varepsilon}/C_{4}$, see Eq. (\ref{eq:F}),
of the long-range interaction for dielectric gratings as compared to ideal
metals (semi-logarithmic scale). For typical grating materials the static
dielectric constant $\varepsilon\left(  0\right)  $ is less than 4, leading to
a reduction below one half.}%
\label{fig:F}%
\end{center}
\end{figure}

However, at large distances retardation effects may become important. They
show up if the separation $\Delta$ between the particle and the grating wall
is comparable to the wavelength corresponding to those virtual transitions in
the particle that contribute with a large oscillator strength. In the case of
an ideal metal the potential is described by the Casimir-Polder formula
\cite{Casimir1948a}. For large distances it has the asymptotic form
$U(\Delta)=-C_{4}/\Delta^{4}$ with the constant
\begin{equation}
C_{4}=\frac{3\hbar c}{8\pi}\,\alpha(0) \label{eq:C4}%
\end{equation}
given by the static polarizability $\alpha(0)$ of the particle. The case of
more realistic grating materials and arbitrary distances is covered by the
theory of Wylie and Sipe \cite{Wylie1984a,Wylie1985a}. It shows that in the
case of \emph{real} metals the asymptotic form of the interaction potential
does not depend on the metal and is identical to the ideal case. For
dielectrics the limiting form depends also on the fourth power of the
distance, but with a reduced interaction constant,
\begin{equation}
C_{4}^{\epsilon}=C_{4}\,\int_{0}^{\infty}\frac{(1+2u^{2})r_{\mathrm{p}%
}(u)-r_{\mathrm{s}}(u)}{(1+u^{2})^{5/2}}\,\frac{u}{2}\,\mathrm{d}u.
\label{eq:F}%
\end{equation}
The reduction depends on the static dielectric constant $\epsilon(0)$ via the
Fresnel coefficients
\begin{equation}
r_{\mathrm{p}}(u)=\frac{\sqrt{1+u^{2}}-\sqrt{\epsilon(0)+u^{2}}}{\sqrt
{1+u^{2}}+\sqrt{\epsilon(0)+u^{2}}}%
\end{equation}
{and}
\begin{equation}
r_{\mathrm{s}}(u)=\frac{\epsilon(0)\sqrt{1+u^{2}}-\sqrt{\epsilon(0)+u^{2}}%
}{\epsilon(0)\sqrt{1+u^{2}}+\sqrt{\epsilon(0)+u^{2}}}\;.
\end{equation}
Figure \ref{fig:F} gives the value of the dielectric reduction factor for
$1\leq\epsilon(0)\leq100$.%

Whether the exact position dependence of the retarded force must be used
depends on the physical situation in the particular interferometric setup. In
most experiments realized so far it was sufficient to use either the static
van der Waals interaction (\ref{eq:C3Lifshitz}) \cite{Bruehl2002a}or the long
range limit of the Casimir-Polder force, Equation~(\ref{eq:C4}).%

\begin{figure}
[ptb]
\begin{center}
\includegraphics[
width=\columnwidth
]%
{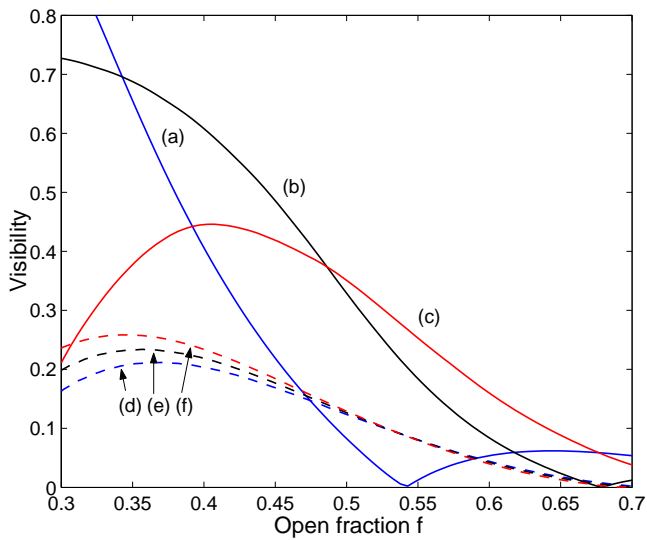}%
\caption{Talbot-Lau visibilities for gratings with a van der Waals interaction
as a function of the open fraction. (a) $L=L_{\lambda}$ (b) $L=0.9L_{\lambda}$
(c) $L=0.8L_{\lambda}$. The corresponding classical visibilities are
given by the
curves (d), (e), (f), respectively.}%
\label{fig:C3visib}%
\end{center}
\end{figure}

Figure~\ref{fig:C3visib} shows the typical effect of a finite grating
interaction, and should be compared to the results for the ideal grating in
Fig.~\ref{fig:idealvisib}. Here we assume a particle with mass
$1000\,\mathrm{amu}$, a van der Waals interaction with $C_{3}%
=10\,\mathrm{meV\,nm^{6},}$ and we take gratings with a period of
$d=1\,\mathrm{\mu m}$ and a thickness of $b=0.2\,\mathrm{\mu m}$ separated by
a distance of $L=0.2\,$m. One observes that the expected quantum visibilities
deviate noticeably from the ideal case. Moreover, the classical expectations
(given by the dashed lines) differ completely from the ideal expectation and
display now a weak velocity dependence. At an open fraction of $0.5$ they now
yield a finite contrast amounting to $12.4\%-12.9\%$ for the three settings.
The respective quantum expectations are also larger than the in the force-free
calculation. (They increased from 14.7\% to 32.9\% at $L/{L_{\lambda}}=0.9$
and from 25.4\% to 35\% at $L/{L_{\lambda}}=0.8$.) This is a typical
phenomenon. The attractive force tends to act as if the open fraction of the
grating was decreased.

\subsection{Gratings of light}

It is clear from Equations (\ref{eq:wxgeneral}) and (\ref{eq:qsignal1}) that
the first and the third gratings in the Talbot-Lau setup must be absorptive to
generate an observable contrast pattern. However, as discussed in
\cite{Brezger2003a} the second grating may be a pure phase grating as well.
Such a \emph{mixed} interferometer can be realized by the off-resonant
interaction with a standing light wave \cite{Kapitza1933a}, see
\cite{Nairz2001a} for the laser diffraction of large molecules.

If we take a TEM$_{00}$ mode of wavelength $\lambda_{\mathrm{L}}%
=2\pi/k_{\mathrm{L}}$ and waist $w$ produced by a laser of power $P_{\text{L}%
}$ then the dipole force leads to the phase shift
\begin{equation}
\tilde{t}(x)=\exp\left(  \mathrm{i}\sqrt{2\pi}\frac{8P_{\text{L}}%
\alpha_{\omega}}{\hbar cv_{z}w}\cos^{2}(k_{\mathrm{L}}x)\right)  ,
\label{eq:laserphase}%
\end{equation}
as follows from an integration over the gaussian beam at central passage. Here
$\alpha_{\omega}$ is the scalar polarizability [73]
of the particle at Laser frequency $\omega_{\mathrm{L}}=ck_{\mathrm{L}}$, where
we assume that $\operatorname{Im}(\alpha_{\omega})=0$ so that photon
absorption can be neglected. The classical momentum kick in (\ref{eq:Tcltilde}%
) that corresponds to the dipole force reads
\begin{equation}
Q(x)=\sqrt{2\pi}\frac{8P_{\text{L}}\alpha_{\omega}\,k_{\mathrm{L}}}{cv_{z}%
w}\sin(2k_{\mathrm{L}}x)\;. \label{eq:laserkick}%
\end{equation}
Using these expressions in (\ref{eq:qeikonal}) and (\ref{eq:cdef}) one obtains
the predictions for the quantum and classical density patterns produced by a
standing light wave in the same way as with material gratings.

In the present section we showed that the Wigner description of the Talbot-Lau
effect permits the effects of the grating interaction to be incorporated
easily, in terms of a simple modification of the Fourier coefficients. As
discussed in the next section the effect of decoherence can be similarly
incorporated into the formalism.

\section{Accounting for decoherence}

\label{sec:deco}Having formulated the Talbot-Lau dynamics in the Wigner
representation it is now easy to include the effects of decoherence. More
specifically, we consider the markovian interaction of the interfering
particle with other, unobserved degrees of freedom (the environment)
\cite{Joos2003a,Zurek2003a}. The resulting formation of quantum correlations
(or \emph{entanglement}) between the particle and the environment leads to a
loss of coherence in the particle state that may be understood from the fact
that a measurement of the environmental degree of freedom could reveal
(partial) which-way information on the particle's whereabouts.

We note that a number of studies have been undertaken recently that describe a
loss of visibility in matter wave interference
\cite{Bonifacio2002a,Facchi2002a,Viale2003a}. Here we focus on the Talbot-Lau
effect and on a formulation that is sufficiently realistic to permit
quantitative predictions about experiments with mesoscopic bodies
\cite{Hornberger2003a,Hackermuller2004a}.

Two important decoherence mechanisms for large, interfering particles are
collisions with background gas particles and the thermal emission of
electromagnetic radiation. Both effects may be treated in the Markov
approximation, which implies that the effect of the environmental coupling can
be described by independent, separate events (such as the emission of a photon
or the collision with a gas particle).

\subsection{The effect of a single decoherence event}

The change in the state of the interfering particle due to a single event can
be obtained by performing a partial trace over the entangled state with
respect to the unobserved degrees of freedom. For particles with a large mass
and for the decoherence mechanisms considered in this article the density
matrix in position representation changes just by a multiplication,
\begin{equation}
\varrho^{\prime}(\mathbf{R}_{1},\mathbf{R}_{2})=\varrho(\mathbf{R}%
_{1},\mathbf{R}_{2})\,\eta(\mathbf{R}_{1}-\mathbf{R}_{2}). \label{eq:decoeff}%
\end{equation}
The factor $\eta(\mathbf{R}_{1}-\mathbf{R}_{2}),$ which may be called the
decoherence function, describes the decay of the off-diagonal elements (the
coherences) of $\hat{\varrho}$ due to a single event. In
Sect.~\ref{sec:decofunction} we will derive the form of realistic decoherence
functions for the most important decoherence mechanisms. For the time being it
is sufficient to note that the conservation of the trace in (\ref{eq:decoeff})
ensures that
\begin{equation}
\lim_{\mathbf{\mathbf{R}_{1}\rightarrow\mathbf{R}_{2}}}\eta(\mathbf{R}%
_{1}-\mathbf{R}_{2})=1, \label{eq:ondiag}%
\end{equation}
so that the diagonal elements of the state are unchanged by (\ref{eq:decoeff}%
). Moreover, the hermiticity of $\hat{\varrho}$ implies $\eta\left(
-\mathbf{R}\right)  =\eta^{\ast}\left(  \mathbf{R}\right)  ,$ and from the
fact that the purity cannot be increased by a partial trace it follows that
$|\eta(\mathbf{R}_{1}-\mathbf{R}_{2})|\leq1.$

If the state is expressed in terms of the Wigner function its change
(\ref{eq:decoeff}) reads%

\begin{equation}
W^{\prime}(\mathbf{R},\mathbf{P})=\int\mathrm{d}\mathbf{Q}\,\bar{\eta
}(\mathbf{Q})W(\mathbf{R},\mathbf{P}-\mathbf{Q}). \label{Wconv}%
\end{equation}
with $\bar{\eta}(\mathbf{Q})$ the Fourier transform of the decoherence
function,%
\begin{equation}
\bar{\eta}(\mathbf{Q})=\frac{1}{\left(  2\pi\hbar\right)  ^{3}}\int
\mathrm{d}\mathbf{R~}e^{-\mathrm{i}\mathbf{QR/\hbar}}\eta\left(
\mathbf{R}\right)  .
\end{equation}
Clearly, the effect of a decoherence event on the Wigner function is to smear
it out in the momentum direction.

As discussed in Sect.~\ref{sec:lcoh} the coherently evolving, stationary state
of the beam in a Talbot-Lau interferometer is described by the function%
\begin{equation}
W_{\mathrm{beam}}(\mathbf{r}+z\mathbf{e}_{z},\mathbf{p}+p_{z}\mathbf{e}%
_{z})=w(z;\mathbf{r},\mathbf{p})g(p_{z}).\label{Wbeam}%
\end{equation}
In a typical setup the grating constant and the grating separation differ by
six orders of magnitude so that $\mathbf{p}$ varies on a scale in
(\ref{Wbeam}) that is much smaller than the magnitude of $p_{z}.$ Our basic
approximation is now to assume that the width of $\bar{\eta}$ is small
compared to the scale over which $W_{\mathrm{beam}}$ varies in $p_{z}$. This
assumption is particularly unproblematic in the Talbot-Lau setup, where the
sensitivity to changes in the longitudinal momentum is rather weak. It follows
that the new state of the transverse coordinates is approximately given by
integrating the full state with respect to the longitudinal momentum,%

\begin{equation}
w^{\prime}(z;\mathbf{r},\mathbf{p})\simeq\int\mathrm{d}p_{z}^{\prime
}\mathbf{~}W_{\text{beam}}^{\prime}\left(  \mathbf{r}+z\mathbf{e}%
_{z},\mathbf{p}+p_{z}^{\prime}\mathbf{e}_{z}\right)  . \label{basicapprox}%
\end{equation}
Inserting (\ref{Wconv}) and (\ref{Wbeam}) yields the change in the state of
the transverse coordinates%

\begin{equation}
w^{\prime}(z;\mathbf{r},\mathbf{p})=\int\mathrm{d}\mathbf{q~}\bar{\eta
}_{\mathrm{2d}}(\mathbf{q)~}w\left(  z;\mathbf{r},\mathbf{p}-\mathbf{q}%
\right)  \label{wdecotrans}%
\end{equation}
with%
\begin{equation}
\bar{\eta}_{\mathrm{2d}}(\mathbf{q)}\equiv\int\mathrm{d}q_{z}~\bar{\eta
}(\mathbf{q}+q_{z}\mathbf{e}_{z}). \label{etabarzwod}%
\end{equation}
It follows from (\ref{etabarzwod}) that in the position representation of the
transverse state, $\rho(\mathbf{r}_{1},\mathbf{r}_{2}),$ the decoherence
function enters without modification,%

\begin{align}
\rho^{\prime}(\mathbf{r}_{1},\mathbf{r}_{2})  &  =\rho(\mathbf{r}%
_{1},\mathbf{r}_{2})\,\int\mathrm{d}\mathbf{q~}\mathrm{e}^{\mathrm{i}%
\mathbf{q}\left(  \mathbf{r}_{1}-\mathbf{r}_{2}\right)  /\hbar}\bar{\eta
}_{\mathrm{2d}}(\mathbf{q)}\nonumber\\
&  =\rho(\mathbf{r}_{1},\mathbf{r}_{2})\,\eta\left(  \mathbf{r}_{1}%
-\mathbf{r}_{2}\right)  ,
\end{align}
but restricted to the $xy$-plane, $\eta\left(  \mathbf{r}\right)  \equiv
\eta\left(  \mathbf{r}+0\mathbf{e}_{z}\right)  .$

Using the Wigner function it is now possible to evaluate the effect of a
\emph{single} decoherence event that takes place at a distance $z$ behind the
first grating. One merely propagates the state to the longitudinal position
$z$ using the coherent transformations (\ref{eq:freetrafo1}) and
(\ref{eq:gratingtrafo2x}), then applies (\ref{wdecotrans}), and propagates the
state over the remaining distance to the third grating. Within the paraxial
approximation no additional error is introduced by this procedure since the
composition property of the evolution holds exactly. The result takes a simple
form (both for $z<L$ and for $z>L$) once the momentum is integrated to yield
the position density $w\left(  \mathbf{r}\right)  $ in front of the third
grating. It is given by an integral of the form (\ref{eq:wxgeneral}),
\begin{equation}
w\left(  \mathbf{r}\right)  =\int\mathrm{d}\mathbf{r}_{1}|t_{1}\left(
\mathbf{r}_{1}\right)  |^{2}\hat{h}_{z}(\mathbf{r};\mathbf{r}_{1}),
\end{equation}
where the coherent kernel $h\left(  \mathbf{r},\mathbf{r}_{1}\right)  $ from
(\ref{eq:hdef}) is replaced by
\begin{equation}
\hat{h}_{z}(\mathbf{r};\mathbf{r}_{1})=\int\mathrm{d}\mathbf{q}\,\bar{\eta
}_{\mathrm{2d}}(\mathbf{q})~h\left(  \mathbf{r},\mathbf{r}_{1}-\frac
{L-|z-L|}{p_{z}}\mathbf{q}\right)  .\label{eq:hhat}%
\end{equation}
This shows clearly that close to the first and to the third grating, at $z=0$
and $z=2L$, a decoherence event will not affect the interference pattern
while, for monotonically decreasing $\bar{\eta}_{\mathrm{2d}}(\mathbf{q}),$ the
interference is most strongly affected by decoherence events that take place
in the vicinity of the second grating, at $z=L$. This is consistent with the
notion that in the Talbot-Lau setup diffraction takes place only at the second
grating, while the first grating acts as an array of coherence slits.

We take the grating function again to be periodic in $x$ (with period $d)$ and
uniform in $y.$ This means that the discussion can be confined to the
$x$-coordinate like in Sect. \ref{sec:TLsetup}. Using the Fourier
decomposition (\ref{eq:Tfourier}) one finds that the coherent kernel
(\ref{eq:hdef}) reads in the one-dimensional case%
\begin{align}
h\left(  x;x_{1}\right)   &  \equiv\int\mathrm{d}y\mathrm{d}y_{1}h\left(
x\mathbf{e}_{x}+y\mathbf{e}_{y};x_{1}\mathbf{e}_{x}+y_{1}\mathbf{e}_{y}\right)
\label{honed}\\
&  =\frac{p_{z}}{2L\hbar}\sum_{m}\mathrm{\exp}\left(  2\pi\mathrm{i}%
m\frac{x+x_{1}}{2d}\right)  \tilde{B}_{m}^{(\lambda)}.\nonumber
\end{align}
It follows with (\ref{eq:hhat}) that in the presence of a decoherence event
the kernel takes the form%
\begin{align}
\hat{h}_{z}\left(  x;x_{1}\right)  =&\frac{p_{z}}{2L\hbar}\sum_{m}\mathrm{\exp
}\left(  2\pi\mathrm{i}m\frac{x+x_{1}}{2d}\right)  
\nonumber\\
&\times \tilde{B}_{m}^{(\lambda
)}\eta\left(  -m\frac{d}{2}\frac{L-|z-L|}{{L_{\lambda}}}\mathbf{e}_{x}\right)
,
\end{align}
where the three dimensional decoherence function from (\ref{eq:decoeff})
enters with its dependence along the $x$-axis. The comparison with
(\ref{honed}) shows that the modified interference pattern corresponding to a
single decoherence event at position $z$ is completely described by a
modification of the coherent Fourier components (\ref{eq:Btilde})
\begin{equation}
\widetilde{B}_{m}^{(\lambda)}\rightarrow\widetilde{B}_{m}^{(\lambda)}%
\,\eta\left(  -m\frac{d}{2}\frac{L-|z-L|}{{L_{\lambda}}}\mathbf{e}_{x}\right)
. \label{eq:Bdeco}%
\end{equation}

\subsection{An alternative to the master equation}

One can now account for probabalistically occurring decoherence events by
considering the change in the final interference pattern due to events that
occur with rate $R(z)$ in the interval $(z;z+\mathrm{d}z)$. It follows from
(\ref{eq:Bdeco}) that the corresponding Fourier coefficients satisfy the
differential equation%

\begin{equation}
\frac{\mathrm{d}}{\mathrm{d}z}\widehat{B}_{m}^{(\lambda)}=R(z)\left[
\widehat{B}_{m}^{(\lambda)}\,\eta\left(  -m\frac{d}{2}\frac{L-|z-L|}%
{{L_{\lambda}}}\mathbf{e}_{x}\right)  -\widehat{B}_{m}^{(\lambda)}\right]  .
\label{Bdiffeq}%
\end{equation}
It describes the change of the interference pattern with an increasing size of
the interval where decoherence events may occur. The integration of
(\ref{Bdiffeq}) over the whole range $z\in\left(  0;2L\right)  $ of admitted
decoherence then yields the coefficients characterizing the modified pattern.
They are given by%
\begin{widetext}
\begin{equation}
\widehat{B}_{m}^{(\lambda)}=\widetilde{B}_{m}^{(\lambda)}\exp\left(  -\int
_{0}^{2L}R(z)\left[  1-\eta\left(  -m\frac{d}{2}\frac{L-|z-L|}{{L_{\lambda}}%
}\mathbf{e}_{x}\right)  \right]  \mathrm{d}z\right)  , \label{eq:Bhat}%
\end{equation}
\end{widetext}
with $\widetilde{B}_{m}^{(\lambda)}$ the coefficients of the coherent
evolution. This is the central result of this section. It shows that the
effects of markovian decoherence of the form (\ref{eq:decoeff}) can be
calculated analytically if the setup is insensitive to longitudinal
correlations as in the Talbot-Lau interferometer. It follows immediately that
the position density and the visibility of the modulation signal are given by
the formulas (\ref{eq:qdensity}) and (\ref{eq:qvisibility}), respectively, if
the coherent coefficients $\widetilde{B}_{m}^{(\lambda)}$ are replaced by
those of the incoherent evolution (\ref{eq:Bhat}).

The result (\ref{eq:Bhat}) can be easily generalized to the asymmetric
Talbot-Lau interferometer. The case of several independent decoherence
mechanisms is also easily incorporated. The resulting interference pattern is
then characterized by a product of the corresponding exponentials in
(\ref{eq:Bhat}).

It is important to note that the basic Fourier components $m=0$ are not
affected by decoherence, since $\eta\left(  0\right)  =1.$ This shows that the
mean count rate does not change due to the presence of decoherence, as is to
be expected from the conservation of the norm in (\ref{eq:decoeff}). The
reduction of the observed visibility assumes a compact form if the modulation
signal (\ref{eq:qsignal2}) is (approximately) sinusoidal, as is typically the
case for gratings with an open fraction of $f\simeq0.5$. Then only the
coefficients $\widehat{B}_{0}^{(\lambda)}$and $\widehat{B}_{2}^{(\lambda)}%
$contribute to the visibility if the grating periods are equal, $r=2$. With
$\eta\left(  0\right)  =1$ it follows that the reduced visibility is given by%
\begin{equation}
\mathcal{V}=\mathcal{V}_{0}\exp\left(  -\int_{0}^{2L}R(z)\left[  1-\eta\left(
-d\frac{L-|z-L|}{{L_{\lambda}}}\mathbf{e}_{x}\right)  \right]  \mathrm{d}%
z\right)  ,\label{visibred}%
\end{equation}
where $\mathcal{V}_{0}$ indicates the visibility in the absence of
decoherence. This formula is particularly intuitive if the Talbot criterion is
met, $L=L_{\lambda}$. Then the argument of $\eta$ contains the separation of
two paths that start and end at common points and pass the second grating
through neighboring slits. At $z=L$ it is equal to the grating constant $d,$
which shows that the Talbot-Lau interference with equal gratings is based on
the interference through neighboring\emph{\ }slits. Also at other positions
$z$ a reduced magnitude of $\eta$ suppresses the visibility whenever 
the change in the
environmental state is able to resolve the corresponding path separation.
Higher orders of the Talbot effect, $L=mL_{\lambda}$ with $m\in\mathbb{N}$,
correspond to multiple slit separations $md$. For longitudinal velocities that
deviate from the Talbot criterion with $L\neq mL_{\lambda}$ the argument is
replaced by an \textquotedblleft effective\textquotedblright\ path separation.

It should be emphasized that our derivation of (\ref{eq:Bhat}) and
(\ref{visibred}) is rather different from solving the markovian master
equation corresponding to the decoherence mechanism. In Sect.
\ref{sec:mastereq} we obtain a solution of the master equation corresponding
to decoherence of the type (\ref{eq:decoeff}) for a general interfering state
in the paraxial approximation. An expression analogous to (\ref{visibred}) is
found there, albeit in a time-dependent formulation, see (\ref{finpara}). This
vindicates our approximation (\ref{basicapprox}).

The present formulation has the particular advantage that the \emph{rate} $R$
of decoherence events and their \emph{effect} $\eta$ appear separately in the
equation. This might seem to be a complication, since these two quantities
must be calculated independently by quantum mechanical means. However, they
are often needed with different degrees of sophistication. For example, often
one must take into account the position dependence of the rate. This is easily
incorporated in the present framework, while solving a corresponding master
equation would be incomparably more complicated.

\subsection{Quantum decoherence vs. a classical stochastic process}

Having treated the effect of environmental coupling on the quantum evolution,
we can now turn to its effects in the classical description. In
Sect.~\ref{sec:class} the classical expectation was calculated in terms of the
phase space density $f(\mathbf{r},\mathbf{p}).\ $The close analogy between the
quantum problem and the classical calculation allows to map the Wigner
representation of a decoherence event (\ref{wdecotrans}) to the classical
description. It follows from (\ref{wdecotrans}) that the effect of a
decoherence event can be interpreted on the classical level as a probabilistic
momentum kick,%
\begin{equation}
f^{\prime}(\mathbf{r},\mathbf{p})=\int\mathrm{d}\mathbf{q~}\bar{\eta
}_{\mathrm{2d}}(\mathbf{q)~}f\left(  \mathbf{r},\mathbf{p}-\mathbf{q}\right)
.
\end{equation}
Indeed, the properties $\eta\left(  0\right)  =1$ and $\eta\left(
-\mathbf{r}\right)  =\eta^{\ast}\left(  \mathbf{r}\right)  $ of the
decoherence function imply that $\bar{\eta}_{\mathrm{2d}}$ has the features of
a probability density, $\bar{\eta}_{\mathrm{2d}}(\mathbf{q)}\geq0$ and
$\int\mathrm{d}\mathbf{q~}\bar{\eta}_{\mathrm{2d}}\left(  \mathbf{q}\right)
=1.$ From the close analogy of the classical and the quantum expressions for
the free evolution and the passage through a grating it is easy to see that
the above derivation of the modified pattern holds in the classical
formulation as well if one replaces the quantum coefficients $\widetilde
{B}_{m}^{(\lambda)}$ in (\ref{eq:Bhat}) by their classical counterparts
$\widetilde{B}_{m}^{(0)}$.

From this one might be led to conclude that the decoherence described in
(\ref{eq:decoeff}) was a \textquotedblleft classical\ effect\textquotedblright%
. In our view this would be a misinterpretation, since a probabilistic
formulation is possible only if the Wigner function is non-negative
everywhere, that is, if it cannot be distinguished from a classical
probability distribution. If the Wigner function is negative in some parts, as
is the case for an interfering state, any stochastic interpretation is
invalidated by the occurring flux of a \textquotedblleft negative
probability\textquotedblright. Notwithstanding this, once the motional state
has turned into a classical state without negativities in the Wigner function
the additional loss of visibility in the quantum description is indeed
indistinguishable from a corresponding classical stochastic process.

\section{Realistic decoherence functions}

\label{sec:decofunction}In the following we discuss the form of realistic
decoherence functions that can be used to obtain quantitative predictions on
the effects of decoherence in matter wave experiments. We focus on the most
important mechanisms for large, massive objects, namely, collisions with
particles from the background gas and the emission of heat radiation. We note
that simple estimates of these effects on material particles can be found in
\cite{Joos1985a,Tegmark1993a,Alicki2002a}.

\subsection{Decoherence by collisions with gas particles}

A very important source of decoherence is the unavoidable presence of a
background gas in the experimental apparatus. Typically, the mass of the
interfering particles is much larger than the mass of the gas particles,
$m_{\mathrm{p}}\gg m_{\mathrm{g}},$ and the interaction is of the monopole
type. In this case the decoherence function reads
\cite{Joos1985a,Gallis1990a,Hornberger2003b}
\begin{widetext}
\begin{equation}
\eta(\mathbf{R}_{1},\mathbf{R}_{2})=\operatorname{tr}_{\text{gas}}\left\{
\mathrm{\exp}\left(  -\mathrm{i}\mathbf{\hat{P}}_{\text{gas}}\frac
{\mathbf{R}_{2}}{\hbar}\right)  \hat{S}_{0}^{\dag}~\mathrm{\exp}\left(
\mathrm{i}\mathbf{\hat{P}}_{\text{gas}}\frac{\mathbf{R}_{2}-\mathbf{R}_{1}%
}{\hbar}\right)  \hat{S}_{0}~\mathrm{\exp}\left(  -\mathrm{i}\mathbf{\hat{P}%
}_{\text{gas}}\frac{\mathbf{R}_{1}}{\hbar}\right)  \hat{\varrho}_{\text{gas}%
}\right\}  \; \label{etacol1}%
\end{equation}
where $\mathbf{\hat{P}}_{\text{gas}}$ is the momentum operator of the gas
particles and $\hat{S}_{0}$ the center-of-mass scattering operator. The trace
over the scattered gas particle in (\ref{etacol1}) can be evaluated if it is
in a (thermal) state that is diagonal in momentum and characterized by the
distribution $\mu_{\text{gas}}\left(  \mathbf{P}\right)  .$ One obtains
\cite{Gallis1990a,Hornberger2003b}%
\begin{equation}
\eta(\mathbf{R}_{1},\mathbf{R}_{2})=\int\mathrm{d}\mathbf{P}~\mu_{\text{gas}%
}\left(  \mathbf{P}\right)  \bigg[1-\int\mathrm{d}\mathbf{P}^{\prime}~\left(
1-\mathrm{e}^{\mathrm{i}\left(  \mathbf{P}-\mathbf{P}^{\prime}\right)  \left(
\mathbf{R}_{1}-\mathbf{R}_{2}\right)  /\hbar}\right)  \underset{X_{\Omega}%
}{\underbrace{\frac{\left(  2\pi\hbar\right)  ^{3}}{\Omega}|\langle
\mathbf{P}^{\prime}|\mathcal{\hat{T}}_{0}|\mathbf{P}\rangle|^{2}}%
}~\bigg] \label{tacol2}%
\end{equation}
\end{widetext}
with $\hat{T}_{0}=\mathrm{i}(1-\hat{S}_{0}).$ Awkwardly, the last two
expressions in (\ref{tacol2}), which are indicated by $X_{\Omega}$, involve two
quantities that are arbitrarily large. One is the \textquotedblleft
quantization volume\textquotedblright\ $\Omega$, which originates from the
normalization of the thermal state $\hat{\varrho}_{\text{gas}},$ and the other
is the \emph{square }of the delta function appearing in the matrix element
$\langle\mathbf{P}^{\prime}|\hat{T}_{0}|\mathbf{P}\rangle=f(\mathbf{P}%
^{\prime},\mathbf{P})\delta\left(  P^{\prime}-P\right)  /(2\pi\hbar P).$ Here
$f(\mathbf{P}^{\prime},\mathbf{P})$ is the scattering amplitude (which must
not be confused with the classical phase space density from Sect.
\ref{sec:class}). Since the decoherence function is well defined by Equation
(\ref{etacol1}) these two infinite quantities must cancel if the limit
$\Omega\rightarrow\infty$ is taken properly. As argued in
\cite{Hornberger2003b} physical consistency requirements lead to
\begin{equation}
\lim_{\Omega\rightarrow\infty}X_{\Omega}=\frac{|f(\mathbf{P}^{\prime
},\mathbf{P})|^{2}}{\sigma\left(  P\right)  }\frac{\delta\left(  P^{\prime
}-P\right)  }{P^{2}} \label{replacement}%
\end{equation}
where $\sigma\left(  P\right)  $ is the total scattering cross section%
\begin{equation}
\sigma\left(  P\right)  =\int\!\!\mathrm{d}\mathbf{n~}|f(P\mathbf{n}%
,\mathbf{P})|^{2}.
\end{equation}
With the replacement (\ref{replacement}) one gets%
\begin{align}
\eta(\mathbf{R}_{1},\mathbf{R}_{2})&=\int\!\!\!\mathrm{d}\mathbf{P}~\frac
{\mu_{\text{gas}}\left(  \mathbf{P}\right)  }{\sigma(P)}
\\
&\times\int\!\!\mathrm{d}%
\mathbf{n}~\left\vert f\left(  \cos(\theta)\right)  \right\vert ^{2}%
\,\mathrm{e}^{\mathrm{i}\left(  \mathbf{P}-P\mathbf{n}\right)  \left(
\mathbf{R}_{1}-\mathbf{R}_{2}\right)  /\hbar}.
\nonumber
\end{align}
As already anticipated in Equation (\ref{eq:decoeff}) this function depends
only on the position difference $\mathbf{R}_{1}-\mathbf{R}_{2}.$ For an
isotropic distribution of the gas momenta, $\mu_{\text{gas}}\left(
\mathbf{P}\right)  =\nu_{\text{gas}}\left(  P\right)  /\left(  4\pi
P^{2}\right)  ,$ the expression can be further simplified noting that it
depends only on the distance $\Delta R=|\mathbf{R}_{1}-\mathbf{R}_{2}|.$ One obtains%

\begin{align}
\eta(\Delta R)&=\int_{0}^{\infty}\!\!\!\mathrm{d}P\;\frac{\nu_{\text{gas}}%
(P)}{\sigma(P)} \label{etacoll}
\\
&\times\int\!\!\mathrm{d}\Omega~\left\vert f\left(  \cos
(\theta)\right)  \right\vert ^{2}\,{\mathrm{sinc}}\!\left(  {\sin\!\left(
\frac{\theta}{2}\right)  }\frac{2P\Delta R}{\hbar}\right) %
\nonumber
\end{align}
with $\operatorname{sinc}\left(  x\right)  =\sin\left(  x\right)  /x.$ The
argument of the $\operatorname{sinc}$ function is equal to the momentum
transfer during the collision times the distance $\Delta R$ in units of
$\hbar.$ This indicates that whenever the change in the state of the gas
particle suffices to resolve\ the distance $\Delta R$ the corresponding
coherences in the motional state will be suppressed.

Let us turn to the second ingredient to the decoherence formula
(\ref{visibred}), the scattering rate $R(z).$ It is usually expressed in terms
of an effective cross section, $R\left(  z\right)  =n\left(  z\right)
\sigma_{\mathrm{eff}},$ with $n\left(  z\right)  $ the number density of the
background gas. For a constant density, and again $m_{\mathrm{p}}\gg
m_{\mathrm{g}},$ the effective cross section depends only on the velocity of
the interfering particle. It is given by
\begin{equation}
\sigma_{\mathrm{eff}}\left(  v_{\mathrm{p}}\right)  =\int\mathrm{d}%
\mathbf{P}~\mu_{\text{gas}}\left(  \mathbf{P}\right)  \sigma\left(
|\mathbf{P}-m_{\mathrm{g}}v_{\mathrm{p}}\mathbf{e}_{\mathrm{z}}|\right)
\frac{|\mathbf{P/}m_{\mathrm{g}}-v_{\mathrm{p}}\mathbf{e}_{\mathrm{z}}%
|}{v_{\mathrm{p}}}, \label{sigmaeff1}%
\end{equation}
as follows from the derivation of the Boltzmann equation.

The most prominent interaction encountered in molecular scattering is the van
der Waals force between polarizable molecules. At the typical velocities in
matter wave interferometry the scattering depends only on the long-range part
of the interaction potential, $U\left(  r\right)  =-C_{6}/r^{6}$, which is
characterized by a single interaction constant $C_{6}.$ The total cross
section is then independent of mass and given by \cite{Maitland1981a}%
\begin{equation}
\sigma\left(  m_{\mathrm{g}}v\right)  =\frac{\pi^{2}}{\Gamma\left(
2/5\right)  \sin\left(  \pi/5\right)  }\left(  \frac{3\pi}{8}\frac{C_{6}%
}{\hbar v}\right)  ^{2/5}.
\end{equation}
The integration in (\ref{sigmaeff1}) can be done assuming a thermal
distribution of the gas particles$.$ The exact expression is given by a
confluent hypergeometric function, as shown recently by Vacchini
\cite{Vacchini2004a}. Here we note the asymptotic form of the effective cross
section (\ref{sigmaeff1}) for small velocities of the interfering particle. It
reads
\begin{align}
\sigma_{\mathrm{eff}}\left(  v_{\mathrm{p}}\right)  &=\frac{4\pi\Gamma\left(
9/10\right)  }{5\sin\left(  \pi/5\right)  }\left(  \frac{3\pi C_{6}}{2\hbar
}\right)  ^{2/5}\frac{\tilde{v}_{\mathrm{g}}^{3/5}}{v_{\mathrm{p}}}
\nonumber\\
&\times
\left\{
1+\frac{1}{5}\left(  \frac{v_{\mathrm{p}}}{\tilde{v}_{\mathrm{g}}}\right)
^{2}+\mathcal{O}\left(  \frac{v_{\mathrm{p}}}{\tilde{v}_{\mathrm{g}}}\right)
^{4}\right\}  \label{sigmaeff2}%
\end{align}
with $\tilde{v}_{\mathrm{g}}=\left(  2k_{\mathrm{B}}T/m_{\mathrm{g}}\right)
^{1/2}$ the most probable velocity in the gas.

In principle, the interaction constant is given by the Casimir-Polder
expression%
\begin{equation}
C_{6}=\frac{3\hbar}{\pi}\int\mathrm{d}\omega~\alpha_{\mathrm{g}}\left(
\mathrm{i}\omega\right)  \alpha_{\mathrm{p}}\left(  \mathrm{i}\omega\right)
\end{equation}
involving the frequency dependent polarizabilities of the two particles [73].
However, often only the static polarizabilities are
available for larger molecules. In this case a fairly accurate estimate can be
obtained from the Slater-Kirkwood expression \cite{Mason1967a}%
\begin{equation}
C_{6}\simeq\frac{3}{2}\frac{e\hbar}{\sqrt{4\pi\varepsilon_{0}m_{e}}}%
\frac{\alpha_{\mathrm{g}}\left(  0\right)  \alpha_{\mathrm{p}}\left(
0\right)  }{\sqrt{\alpha_{\mathrm{g}}\left(  0\right)  /N_{\mathrm{g}}}%
+\sqrt{\alpha_{\mathrm{p}}\left(  0\right)  /N_{\mathrm{p}}}},
\end{equation}
where $N_{\mathrm{g}}$ and $N_{\mathrm{p}}$ are the number of valence
electrons of the gas molecules and the interfering particle, respectively.

Let us stress again that in the present treatment the effect of a single
collision (\ref{etacoll}) and the rate (\ref{sigmaeff2}) are calculated
separately, which is particularly useful if the two are needed at different
degrees of accuracy. This was the case in the recent experiments on
collisional decoherence \cite{Hornberger2003a,Hackermuller2003b} where the
localization took place on a scale that is by orders of magnitude smaller than
the path separation. Consequently, $\eta$ could be replaced by a simple
Kronnecker-like function in (\ref{eq:Bhat}), while the finite velocity of the
interfering particle within the thermal gas had to be taken into account
properly. Corresponding master equations, based on the microscopic description
of the scattering process, are a subject of current research
\cite{Joos1985a,Gallis1990a,Diosi1995a,Vacchini2001a,Vacchini2001b,Dodd2003a,Hornberger2003b}%
. Although some of those are sufficiently detailed to describe the emergence
of an effective scattering rate (\ref{sigmaeff2}), their application to a
description of the experiment would have been considerably more complicated
than in the present treatment.

\subsection{Decoherence by thermal emission of radiation}

A second decoherence mechanism that is common to all macroscopic objects is
the emission of heat radiation. It starts to play a role in matter wave
interference if one considers macro-molecules or mesoscopic particles. Due to
the large number of internal degrees of freedom, a thermodynamic description
of the distribution of the internal energy is unavoidable. Moreover, their
coupling to the electromagnetic field is quasi-continuous. In general, the
thermally emitted photons will reveal (partial) which-way information on the
whereabouts of the interfering particle and thus lead to decoherence.

We assume that the emission is isotropic, and that the walls of the apparatus,
which absorb an emitted photon, are located in the far field where the
photon's spatial detection probability is given by its momentum distribution.
The conservation of the total momentum then suffices to determine the
transformation of the particle's density operator that would be obtained from
a partial trace over the entangled state between photon and particle. It
follows that the change of the particle center of mass coordinate due to a
single emissions given by%

\begin{equation}
\hat{\varrho}\rightarrow\hat{\varrho}^{\prime}=\int\mathrm{d}\mathbf{k}%
\,\frac{p_{k}(k)}{4\pi k^{2}}\,\hat{U}_{\mathbf{k}}\hat{\varrho}\hat
{U}_{\mathbf{k}}^{\dagger},\label{eq:photonemission}%
\end{equation}
where $p_{k}(k)$ is the probability distribution for photons with wave number
$k=|\mathbf{k}|$ and $\hat{U}_{\mathbf{k}}=\exp(\mathrm{i}\hat{\mathbf{R}%
}\mathbf{k})$ are the momentum translation operators. Note that it is not
necessary to consider the change of the internal degrees of freedom of the
particle, since their state does not get entangled with the center of mass;
this would result only if the emission probability were position dependent.

In position representation, $\varrho(\mathbf{R}_{1},\mathbf{R}_{2}%
)=\langle\mathbf{R}_{1}|\varrho|\mathbf{R}_{2}\rangle$, the transformation
(\ref{eq:photonemission}) reduces the off-diagonal elements of the
center-of-mass state
\begin{equation}
\varrho^{\prime}(\mathbf{R}_{1},\mathbf{R}_{2})=\varrho(\mathbf{R}%
_{1},\mathbf{R}_{2})\eta(\mathbf{R}_{1}-\mathbf{R}_{2}).
\end{equation}
The corresponding decoherence function reads
\begin{equation}
\eta(\Delta R)=\frac{1}{R_{\mathrm{tot}}}\int_{0}^{\infty}\!\!\!\mathrm{d}%
\lambda\,R_{\lambda}(\lambda)\,{\mathrm{sinc}}\left(  2\pi\frac{\Delta
R}{\lambda}\right)  ,
\end{equation}
where the probability distribution $p_{k}$ was expressed in terms of the
spectral photon emission rate,%
\begin{equation}
R_{\lambda}(\lambda)=\frac{2\pi R_{\mathrm{tot}}}{\lambda^{2}}p_{k}\left(
\frac{2\pi}{\lambda}\right)  ,
\end{equation}
and the total photon emission rate
\begin{equation}
R_{\mathrm{tot}}=\int_{0}^{\infty}R_{\lambda}(\lambda)\,\mathrm{d}\lambda.
\end{equation}
In the expression for the Fourier components (\ref{eq:Bhat}) the total rate of
decoherence events $R_{\mathrm{tot}}$ cancels because it gets multiplied by
$\eta.$ One obtains
\begin{widetext}
\[
\widehat{B}_{m}^{(\lambda)}=\widetilde{B}_{m}^{(\lambda)}\exp\left(  -\frac
{1}{v_{z}}\int_{0}^{2L}\mathrm{d}z\int_{0}^{\infty}\mathrm{d}\lambda
~R_{\lambda}(\lambda)\left[  1-\operatorname{sinc}\left(  m\pi\frac{d}%
{\lambda}\frac{L-|z-L|}{{L_{\lambda}}}\right)  \right]  \right)  .
\]
\end{widetext}
This shows clearly how the fringe pattern gets blurred by heat radiation if it
contains photons that have a sufficiently small wavelength to resolve the path
separation. The properties of the interfering particle enter only through the
spectral emission rate $R_{\lambda}=R_{\omega}\times|\mathrm{d}\omega
/\mathrm{d}\lambda|.$

For mesoscopic particles the spectral emission rate deviates from the Planck
law of a macroscopic black body for a number of reasons. First, the photon
wavelengths are typically much larger than the radiating particle, which turns
it into a colored emitter. The density of available transition matrix elements
can be related to the absorption cross section \cite{Friedrich1998a}. Second,
at internal energies where thermal emission is relevant the particle is
usually not in thermal equilibrium with the radiation field, so that there is
no induced emission. Third, the particle is not in contact with a heat bath,
but the emission takes place at a fixed internal energy $E$. Similarly to
Einstein's derivation of the Planck law, these points lead to the expression
\cite{Hansen1998a}%

\begin{equation}
R_{\omega}\left(  \omega\right)  \mathrm{d}\omega=\frac{\omega^{2}}{\pi
^{2}c^{2}}\sigma_{\mathrm{abs}}\left(  E-\hbar\omega;\omega\right)
\frac{d\left(  E-\hbar\omega\right)  }{d\left(  E\right)  }\mathrm{d}\omega.
\label{Romega}%
\end{equation}
The first term is proportional to the mode density. The mean oscillator
strength is described by the absorption cross section at frequency $\omega$
and internal energy $E-\hbar\omega,$ and the ratio of the densities of state
$d\left(  E\right)  $ yields the statistical factor under a strong mixing
assumption. The mean densities of states can be related to the thermodynamic
properties of the particle by a stationary phase evaluation of the inverse
Laplace transform of its partition function \cite{Hoare1970a,Frauendorf1995a}.
This yields $d\left(  E\right)  \sim\exp\left(  S\left(  E\right)
/k_{\mathrm{B}}\right)  $ and therefore%
\begin{equation}
\frac{d\left(  E-\hbar\omega\right)  }{d\left(  E\right)  }\simeq\exp\left[
-\frac{\hbar\omega}{k_{\mathrm{B}}T^{\ast}}-\frac{1}{2C_{V}}\left(
\frac{\hbar\omega}{k_{\mathrm{B}}T^{\ast}}\right)  ^{2}\right]  .
\label{dratio}%
\end{equation}
Here, the internal energy is conveniently expressed in terms of the
micro-canonical temperature
\begin{equation}
T^{\ast}\left(  E\right)  =\left[  \frac{\partial S\left(  E\right)
}{\partial E}\right]  ^{-1},
\end{equation}
with $S\left(  E\right)  $ the entropy. The value of $T^{\ast}$ is equal, up
to small corrections, to that canonical temperature where the mean energy
equals the internal energy. The second term in (\ref{dratio}) contains the
heat capacity $C_{V}$ of the particle. It is the leading correction due to the
finite size of the internal heat bath. This term decreases with increasing
size of the particle and (\ref{dratio}) assumes the canonical form in the
limit $C_{V}\rightarrow\infty$.

With equations (\ref{Romega}) and (\ref{dratio}) one is able to calculate the
temperature dependent spectral emission rate $R_{\lambda}(\lambda,T)$ and the
corresponding decoherence effect. At very small heat capacities the effect of
cooling may have to be taken into account. It can be easily incorporated in
the present framework through a position dependent temperature $T\left(
z\right)  $ determined by the cooling formula%
\begin{equation}
\frac{\mathrm{d}}{\mathrm{d}z}T\left(  z\right)  =-\frac{1}{v_{z}C_{V}}%
\int\hbar\omega R_{\omega}\left(  \omega,T\right)  \mathrm{d}\omega.
\end{equation}

We note that also scattering of photons may lead to decoherence, although room
temperature photons will not limit matter wave interference in the foreseeable
future. The deliberate scattering of a laser beam at interfering atoms at a
resonant cross section was studied in
\cite{Pfau1994a,Clauser1994b,Chapman1995a,Mei2001a,Kokorowski2001a}.

Finally, we emphasize that all the calculations in this paper have been within
the framework of conventional quantum mechanics. However, proposed extensions
of that theory, which produce spontaneous localization of massive particles
due to postulated ``collapse'' terms added to the Schr{\"o}dinger equation, lead to
an evolution of the density operator which mimics decoherence effects
\cite{Bassi2003a}. Hence both the establishment of the framework we develop
here, and accurate models for decoherence mechanisms, are essential to
ascertain whether or not any particular proposed modification of conventional
quantum mechanics can be ruled out by experimental data. We defer such
applications to future articles with a focus on laboratory results.

\section{Equivalence with the master equation}

\label{sec:mastereq}In this final section we show that the procedure to
incorporate decoherence that was used in Sect.~\ref{sec:deco} is equivalent to
solving the corresponding master equation in paraxial approximation. This is
done by identifying the systematic corrections to the paraxial approximation
in terms of the ratio between transverse and longitudinal momenta.

Our starting point is the master equation for a free particle
\begin{align}
\frac{\partial}{\partial t}\hat{\varrho}&=\frac{1}{\mathrm{i}\hbar}\left[
\frac{\mathbf{\hat{P}}^{2}}{2m_{\mathrm{p}}},\hat{\varrho}\right]
\nonumber\\
&-\int\mathrm{d}\mathbf{R\,}\mathrm{d}\mathbf{R}^{\prime}\,\gamma\left(
\mathbf{R-R}^{\prime}\right)  \,\varrho\left(  \mathbf{R,R}^{\prime}\right)
\,\left\vert \mathbf{R}\right\rangle \left\langle \mathbf{R}^{\prime
}\right\vert \label{MasterEqInfMass}%
\end{align}
with localization rate $\gamma.$ It is valid in situations where the
mass of the
particle $m_{\mathrm{p}}$ is sufficiently large so that the effect of the
environmental coupling does not (yet) lead to thermalization. This equation is
usually applicable in interferometric situations where one is interested in
time scales that are much shorter than those of dissipation. In particular, it
describes the effects of scattering of particles with a much smaller mass or
the emission of photons.

It follows from (\ref{MasterEqInfMass}) that the corresponding Wigner function
satisfies%
\begin{align}
\frac{\partial}{\partial t}W\left(  \mathbf{R,P};t\right) & =-\frac{\mathbf{P}%
}{m_{p}}\nabla_{\mathbf{R}}W\left(  \mathbf{R,P};t\right)  
\nonumber\\&-\int
\mathrm{d}\mathbf{P}^{\prime}\mathbf{\,}\bar{\gamma}\left(  \mathbf{P}%
^{\prime}\right)  \,W\left(  \mathbf{R,P}-\mathbf{P}^{\prime};t\right)
\label{MasterforWigner2}%
\end{align}
with $\bar{\gamma}$ the Fourier transform of the localization rate,
\begin{equation}
\bar{\gamma}\left(  \mathbf{P}\right)  =\frac{1}{\left(  2\pi\hbar\right)
^{3}}\int\mathrm{d}\mathbf{R}\,e^{-\mathrm{i}\mathbf{RP}/\hbar}\gamma\left(
\mathbf{R}\right)  .
\end{equation}
Unlike in the previous sections, we describe the motion of the particle by a
Wigner function that is properly \emph{normalized},
\begin{equation}
\int\mathrm{d}\mathbf{R\,}\mathrm{d}\mathbf{P}\,W\left(  \mathbf{R,P}%
;t\right)  =1.
\end{equation}

\subsection{Decoherence of an interfering state}

Now consider the usual scattering situation where the particle enters and
leaves the grating region of an interferometer in a finite period of time, so
that $W\left(  \mathbf{R,P};t\rightarrow\pm\infty\right)  =0$ for all
positions $\mathbf{R}$ of interest. It follows that%
\begin{align}
\int_{-\infty}^{\infty}\mathrm{d}t\,\partial_{t}W\left(  \mathbf{R,P}%
;t\right)  &=W\left(  \mathbf{R,P};+\infty\right)  -W\left(  \mathbf{R,P}%
;-\infty\right)  \nonumber
\\&=0.\label{zerointegral}%
\end{align}
As above, we take the $z$--axis as the longitudinal direction of the
interferometer,%
\begin{align}
\mathbf{R} &  =\mathbf{r}+z\mathbf{e}_{z}~\nonumber\\
\mathbf{P} &  =\mathbf{p}+p_{z}\mathbf{e}_{z},
\end{align}
and denote the transverse positions and momenta by $\mathbf{r}=\left(
x,y\right)  $ and $\mathbf{p}=\left(  p_{x},p_{y}\right)  $, respectively.

At $t=0$ the particle is localized in the region $z<0$ and heading for the
region $z>0$ where decoherence may occur, say, because there is a gas present.
Moreover, we assume that at $t=0$ the particle is already in a nonclassical
motional state, for example because it has just passed a grating. The expected
interference \thinspace pattern is given by the position dependent detection
probability in the $z$-plane which is obtained by integrating the longitudinal
current density over time,%
\begin{equation}
Q\left(  z;\mathbf{r}\right)  =\int\mathrm{d}t\int\mathrm{d}\mathbf{p\,}%
\mathrm{d}p_{z}\,\frac{p_{z}}{m_{p}}W\left(  \mathbf{r+}z\mathbf{e}%
_{z}\mathbf{,p+}p_{z}\mathbf{e}_{z};t\right)  \,.\label{patterndef}%
\end{equation}
In order to compare with the results from Sect.~\ref{sec:deco} we are
ultimately interested in the effect of decoherence on the Fourier transform of
the interference pattern with respect to the transverse coordinates,%
\begin{align}
\bar{Q}\left(  z;\mathbf{q}\right)     =&\frac{1}{\left(  2\pi\hbar\right)
^{2}}\mathbf{\int}\mathrm{d}\mathbf{r\,}e^{-\mathrm{i}\mathbf{qr/}\hbar
}\,Q\left(  z;\mathbf{r}\right)  \nonumber\\
  =&\int\mathrm{d}\mathbf{p\,}\mathrm{d}p_{z}\,\frac{p_{z}}{\left(  2\pi
\hbar\right)  ^{2}m_{p}}\mathbf{\int}\mathrm{d}\mathbf{r\,}e^{-\mathrm{i}%
\mathbf{qr/}\hbar}
\nonumber
\\&\times\int\mathrm{d}t\mathbf{\,}W\left(  \mathbf{r+}%
z\mathbf{e}_{z}\mathbf{,p+}p_{z}\mathbf{e}_{z};t\right)  \nonumber\\
  \equiv&\int\mathrm{d}\mathbf{p\,}\mathrm{d}p_{z}\,\mathbf{\,}S_{a}\left(
z,p_{z};\mathbf{q,p}\right)  .\label{fdef}%
\end{align}
Here we introduced the auxiliary function $S_{a}$. In order to obtain a
differential equation for $S_{a}$ apply $\mathbf{\int}\mathrm{d}%
\mathbf{r\,\exp}\left(  -\mathrm{i}\mathbf{qr/}\hbar\right)  \int
\mathrm{d}t\left[  \cdot\right]  $ to (\ref{MasterforWigner2}). Using
(\ref{zerointegral}) and integrating by parts one finds%
\begin{align}
&\frac{\mathrm{\partial}}{\mathrm{\partial}z}S_{a}\left(  z,p_{z}%
;\mathbf{q,p}\right)     =-\mathrm{i}\frac{\mathbf{q}\,\cdot\mathbf{p}}%
{p_{z}\hbar}S_{a}\left(  z,p_{z};\mathbf{q,p}\right)  \label{STdynamics}\\
&  -m_{p}\int\mathrm{d}p_{z}^{\prime}\mathrm{d}\mathbf{p}^{\prime}\frac
{\bar{\gamma}\left(  \mathbf{p}^{\prime}+p_{z}^{\prime}\mathbf{e}_{z}\right)
\,}{p_{z}-p_{z}^{\prime}}S_{a}\left(  z,p_{z}-p_{z}^{\prime};\mathbf{q,p}%
-\mathbf{p}^{\prime}\right)  \,.\nonumber
\end{align}
In the absence of decoherence this differential equation is immediately
integrated,%
\begin{align}
S_{a}\left(  z,p_{z};\mathbf{q,p}\right)  =&\exp\left(  -\mathrm{i}%
z\frac{\mathbf{q\,}\cdot\mathbf{p}}{p_{z}\hbar}\right)  \,S_{a}\left(
0,p_{z};\mathbf{q,p}\right)  
\\\nonumber &\text{(for }\gamma=0\text{).}%
\end{align}
This decoherence-free solution is used below to obtain a systematic
approximation in the presence of decoherence. But first we introduce the
Fourier transform of $S_{a}$ with respect to the longitudinal momentum
referenced by a fixed characteristic momentum $\bar{p}_{z},$%

\begin{equation}
S_{b}\left(  z,\zeta;\mathbf{q,p}\right)  :=\int\mathrm{d}p_{\delta}%
\,\exp\left(  \mathrm{i}\zeta p_{\delta}/\hbar\right)  \,S_{a}\left(
z,\bar{p}_{z}+p_{\delta};\mathbf{q,p}\right)  \label{SbST}%
\end{equation}
so that%
\begin{equation}
\bar{Q}\left(  z;\mathbf{q}\right)  =\int\mathrm{d}\mathbf{p\,}S_{b}\left(
z,0;\mathbf{q,p}\right)  . \label{ffromSb}%
\end{equation}
The motivation for this definition is that we will assume that $S_{a}\left(
z,p_{z};\mathbf{q,p}\right)  $ is strongly peaked around the characteristic
momentum $p_{z}=\bar{p}_{z}$ and therefore $S_{b}$ should be a slowly varying
function of $\zeta.$ This will form the basis of our approximations below. For
the time being we keep the equations exact.

The dynamics for $S_{b}$ follows from (\ref{STdynamics}).%
\begin{align}
\frac{\mathrm{d}}{\mathrm{d}z}S_{b}\left(  z,\zeta;\mathbf{q,p}\right)
&=\left[  \frac{\mathrm{d}}{\mathrm{d}z}S_{b}\left(  z,\zeta;\mathbf{q,p}%
\right)  \right]  _{\text{coh}.}
\nonumber\\
&+\left[  \frac{\mathrm{d}}{\mathrm{d}z}%
S_{b}\left(  z,\zeta;\mathbf{q,p}\right)  \right]  _{\text{incoh.}%
}\label{SbEOM}%
\end{align}
The coherent part reads%
\begin{align}
&\left[  \frac{\mathrm{d}}{\mathrm{d}z}S_{b}\left(  z,\zeta;\mathbf{q,p}%
\right)  \right]  _{\text{coh}.}=
\\ 
&  -\mathrm{i}\frac{\mathbf{q}%
\cdot\mathbf{p}}{\bar{p}_{z}\hbar}\int\mathrm{d}p_{\delta}\,\exp\left(
\mathrm{i}\zeta p_{\delta}/\hbar\right)  \frac{\bar{p}_{z}}{\bar{p}%
_{z}+p_{\delta}}S_{a}\left(  z,\bar{p}_{z}+p_{\delta};\mathbf{q,p}\right)
\nonumber\\
= &  -\mathrm{i}\frac{\mathbf{q\cdot p}}{\bar{p}_{z}\hbar}S_{b}\left(
z,\zeta;\mathbf{q,p}\right)  \nonumber\\
&  -\mathrm{i}\frac{\mathbf{q}\cdot\mathbf{p}}{\bar{p}_{z}\hbar}\int
\mathrm{d}p_{\delta}\,\exp\left(  \mathrm{i}\zeta\,p_{\delta}/\hbar\right)
\,\mathcal{D}\left(  \frac{p_{\delta}}{\bar{p}_{z}}\right)  S_{a}\left(
z,\bar{p}_{z}+p_{\delta};\mathbf{q,p}\right)  \nonumber
\end{align}
where we used $\bar{p}_{z}/\left(  \bar{p}_{z}+p_{\delta}\right)
=1+\mathcal{D}\left(  p_{\delta}/\bar{p}_{z}\right)  $ with
\begin{equation}
\mathcal{D}\left(  x\right)  :=-\frac{x}{1+x}=\sum_{n=1}^{\infty}\left(
-x\right)  ^{n}\label{Dexpansion}%
\end{equation}
Formally, Equation (\ref{SbST}) allows to introduce a differential operator
for $S_{b},$
\begin{align}
&\left[  \frac{\mathrm{d}}{\mathrm{d}z}S_{b}\left(  z,\zeta;\mathbf{q,p}%
\right)  \right]  _{\text{coh}.}=
\nonumber\\
&-\mathrm{i}\frac{\mathbf{q\cdot p}}{\bar
{p}_{z}\hbar}\left[  1+\mathcal{D}\left(  \frac{-\mathrm{i}\hbar}{\bar{p}_{z}%
}\frac{\mathrm{d}}{\mathrm{d}\zeta}\right)  \right]  S_{b}\left(
z,\zeta;\mathbf{q,p}\right)  .\label{Sbcoh}%
\end{align}
Since $S_{b}$ was constructed to have a weak dependence on $\zeta$ we expect
that the expansion (\ref{Dexpansion}) can be relied on at least in an
asymptotic sense. For the second term in (\ref{SbEOM}) one obtains in a
similar way%

\begin{align}
&\left[  \frac{\mathrm{d}}{\mathrm{d}z}S_{b}\left(  z,\zeta;\mathbf{q,p}%
\right)  \right]  _{\text{incoh}.}=
\nonumber\\
&-\frac{m_{p}}{\bar{p}_{z}}\int 
\mathrm{d}p_{z}^{\prime}\mathrm{d}\mathbf{p}^{\prime}\,\exp\left(
\mathrm{i}\zeta p_{z}^{\prime}/\hbar\right)  \,\bar{\gamma}\left(
\mathbf{p}^{\prime}+p_{z}^{\prime}\mathbf{e}_{z}\right)  \nonumber\\
&  \times\left[  1+\mathcal{D}\left(  \frac{-\mathrm{i}\hbar}{\bar{p}_{z}%
}\frac{\mathrm{d}}{\mathrm{d}\zeta}\right)  \right]  \,S_{b}\left(
z,\zeta;\mathbf{q,p}-\mathbf{p}^{\prime}\right)  .\label{Sbdecoh}%
\end{align}
This integro-differential equation can be further simplified by separating off
the solution of the coherent part (\ref{Sbcoh}) for a vanishing $\zeta
$-dependence and by a Fourier transformation that removes the convolution in
(\ref{Sbdecoh}). This is done by the introduction of a third and final
auxiliary function,%
\begin{align}
S_{c}\left(  z,\zeta;\mathbf{q},\boldsymbol{\rho}\right)  :=&\int
\mathrm{d}\mathbf{p}\exp\left(  \mathrm{i}\frac{\mathbf{q\cdot p}}{\bar{p}%
_{z}\hbar}z-\mathrm{i}\frac{\mathbf{p}\cdot\boldsymbol{\rho}}{\hbar}\right)
\nonumber\\&\times S_{b}\left(  z,\zeta;\mathbf{q,p}\right)  \,.
\end{align}
It reads in terms of the Wigner function%
\begin{align}
S_{c}&\left(  z,\zeta;\mathbf{q},\boldsymbol{\rho}\right)  =   \frac
{1}{\left(  2\pi\hbar\right)  ^{2}}\int\mathrm{d}\mathbf{r\,}\mathrm{d}%
\mathbf{p~}\mathrm{d}p_{z}\mathbf{\,}
\nonumber\\&\times\exp\left(  -\mathrm{i}\frac
{\mathbf{r\cdot q+}\boldsymbol{\rho}\cdot\mathbf{p}}{\hbar}+\mathrm{i}%
\frac{\mathbf{q\cdot p}}{\bar{p}_{z}\hbar}z+\mathrm{i}\zeta\frac{p_{z}-\bar
{p}_{z}}{\hbar}\right)  \label{ScfromWig}\nonumber\\
&  \times\frac{p_{z}}{m_{p}}\int\mathrm{d}t~W\left(  \mathbf{r}+z\mathbf{e}%
_{z}\mathbf{,p}+p_{z}\mathbf{e}_{z};t\right)  .
\end{align}
If one knows the function $S_{c}$ the interference pattern is immediately
obtained since%
\begin{equation}
\bar{Q}\left(  z;\mathbf{q}\right)  =S_{c}\left(  z,0;\mathbf{q},\frac{z}%
{\bar{p}_{z}}\mathbf{q}\right)  \,,\label{ffromSc}%
\end{equation}
as follows from (\ref{ffromSb}). The evolution equation of $S_{c}$ is obtained
from (\ref{SbEOM}). It is now a differential equation,%
\begin{align}
&\frac{\mathrm{d}}{\mathrm{d}z}S_{c}\left(  z,\zeta;\mathbf{q},\boldsymbol{\rho
}\right)  = 
\nonumber\\
&  -\frac{m_{p}}{\bar{p}_{z}}\gamma\left(  \frac{\mathbf{q}}%
{\bar{p}_{z}}z-\boldsymbol{\rho}+\zeta\mathbf{e}_{z}\right)  S_{c}\left(
z,\zeta;\mathbf{q},\boldsymbol{\rho}\right)  \nonumber\\
&  +\frac{\mathbf{q}}{\bar{p}_{z}}\mathcal{D}\left(  \frac{-\mathrm{i}\hbar
}{\bar{p}_{z}}\frac{\mathrm{d}}{\mathrm{d}\zeta}\right)  \nabla
_{\boldsymbol{\rho}}S_{c}\left(  z,\zeta;\mathbf{q},\boldsymbol{\rho}\right)
\nonumber\\
&  -\frac{m_{p}}{\bar{p}_{z}}\gamma\left(  \frac{\mathbf{q}}{\bar{p}_{z}%
}z-\boldsymbol{\rho}+\zeta\mathbf{e}_{z}\right)  \mathcal{D}\left(
\frac{-\mathrm{i}\hbar}{\bar{p}_{z}}\frac{\mathrm{d}}{\mathrm{d}\zeta}\right)
S_{c}\left(  z,\zeta;\mathbf{q},\boldsymbol{\rho}\right)  .\label{PhiEOM}%
\end{align}
In order to find the initial function $S_{c}\left(  0,\zeta;\mathbf{q}%
,\boldsymbol{\rho}\right)  $ we take into account that the initial state is
localized in the left half space,
\begin{equation}
W(\mathbf{r}+z\mathbf{e}_{z},\mathbf{p}+p_{z}\mathbf{e}_{z};0)=0\quad\text{for
}z>0,
\end{equation}
and heading to the right,
\begin{equation}
\lim_{t\rightarrow\infty}W(\mathbf{r}+z\mathbf{e}_{z},\mathbf{p}%
+p_{z}\mathbf{e}_{z};t)=0\quad\text{for }z<0.
\end{equation}
With these conditions the initial function is obtained from (\ref{ScfromWig})
by assuming that the Wigner function evolves freely (without decoherence)
until it reaches the boundary to the decoherence region, $z=0.$ It reads
\begin{align}
&S_{c}\left(  0,\zeta;\mathbf{q},\boldsymbol{\rho}\right)  =   
\frac{1}{\left(  2\pi\hbar\right)  ^{2}}
\nonumber\\&
\times\int\mathrm{d}\mathbf{r\,}\mathrm{d}%
\mathbf{p~}\mathrm{d}p_{z}\mathbf{\,}\exp\left(  -\mathrm{i}\frac
{\mathbf{r\cdot q}+\boldsymbol{\rho}\cdot\mathbf{p}}{\hbar}+\mathrm{i}%
\zeta\frac{p_{z}-\bar{p}_{z}}{\hbar}\right)  \nonumber\\
&  \times\int_{-\infty}^{0}\mathrm{d}z~W\left(  \mathbf{r}+z\mathbf{e}%
_{z}+\frac{z}{p_{z}}\mathbf{p,p}+p_{z}\mathbf{e}_{z};0\right)  .\label{Scnull}%
\end{align}
After solving the differential equation (\ref{PhiEOM}) the resulting, possibly
blurred, interference pattern $Q_{z}\left(  \mathbf{r}\right)  $ is obtained
by taking the inverse Fourier transform of (\ref{ffromSc})%

\begin{equation}
\,Q\left(  z;\mathbf{r}\right)  =\mathbf{\int}\mathrm{d}\mathbf{q\,}%
e^{\mathrm{i}\mathbf{qr/}\hbar}S_{c}\left(  z,0;\mathbf{q},\frac{z}{\bar
{p}_{z}}\mathbf{q}\right)  .
\end{equation}
The evolution equation (\ref{PhiEOM}) shows clearly the hierarchy of
decoherence terms involved in the dynamics. If one neglects the right hand
side of (\ref{PhiEOM}) altogether one obtains the diffraction pattern in the
paraxial approximation,%
\begin{equation}
\bar{Q}_{\text{para}}^{\left(  \gamma=0\right)  }\left(  z;\mathbf{q}\right)
=S_{c}\left(  0,0;\mathbf{q},\frac{z}{\bar{p}_{z}}\mathbf{q}\right)  .
\label{Qbarfree}%
\end{equation}
The first term in (\ref{PhiEOM}) describes the effect of decoherence in the
paraxial approximation, the second term gives the corrections to the
propagation beyond the paraxial approximation, and the third term describes
the modification of the decoherence due to those corrections.

\subsection{Decoherence in the paraxial approximation}

In the simplest approximation we neglect the corrections in (\ref{PhiEOM}) due
to the $\mathcal{D}$-terms. In this case (\ref{PhiEOM}) can be immediately
integrated,%
\begin{align}
S_{c}\left(  z,\zeta;\mathbf{q},\boldsymbol{\rho}\right)  =&\exp\left[
-\frac{m_{p}}{\bar{p}_{z}}\int_{0}^{z}\,\gamma\left(  \frac{z^{\prime
}\mathbf{q}}{\bar{p}_{z}}-\boldsymbol{\rho}+\zeta\mathbf{e}_{z}\right)
\mathrm{d}z^{\prime}\right]  
\nonumber\\
&\times S_{c}\left(  0,\zeta;\mathbf{q},\boldsymbol{\rho
}\right)  .
\end{align}
It follows from (\ref{ffromSc}) that the resulting interference pattern is
characterized by%

\begin{align}
\bar{Q}_{\text{para}}^{\left(  \gamma\neq0\right)  }\left(  z;\mathbf{q}%
\right)     =&
\exp\left[  -\frac{m_{p}}{\bar{p}_{z}}\int_{0}^{z}%
\,\gamma\left(  \frac{\left(  z^{\prime}-z\right)  \mathbf{q}}{\bar{p}_{z}%
}\right)  \mathrm{d}z^{\prime}\right]
\nonumber\\
& \times \bar{Q}_{\text{para}}^{\left(
\gamma=0\right)  }\left(  z;\mathbf{q}\right) \nonumber\\
  =&\exp\left[  -\int_{0}^{t}\,\gamma\left(  \frac{\left(  t^{\prime
}-t\right)  \mathbf{q}}{m_{p}}\right)  \mathrm{d}t^{\prime}\right]
  \bar
\nonumber\\
& \times
{Q}_{\text{para}}^{\left(  \gamma=0\right)  }\left(  z;\mathbf{q}\right)
\label{finpara}%
\end{align}
with $t:=zm_{p}/\bar{p}_{z}.$ Using (\ref{Qbarfree}) and (\ref{patterndef}) we
obtain the final pattern corresponding to a solution of the master equation
(\ref{MasterEqInfMass}) in paraxial approximation.
\begin{widetext}
\begin{align}
Q_{\text{para}}^{\left(  \gamma\neq0\right)  }\left(  z;\mathbf{r}\right)   &
=\int\mathrm{d}\mathbf{r}^{\prime}\mathrm{d}\mathbf{p}\mathrm{d}%
\mathbf{q}\frac{1}{\left(  2\pi\hbar\right)  ^{2}}\exp\left(  \mathrm{i}%
\left(  \mathbf{r}-\mathbf{r}^{\prime}-\frac{z}{\bar{p}_{z}}\mathbf{p}\right)
\frac{\mathbf{q}}{\hbar}\right) \exp\left(  -\frac{m_{p}}{\bar{p}_{z}}\int_{0}^{z}\gamma\left(
\frac{z-z^{\prime}}{\bar{p}_{z}}\mathbf{q}\right)  \mathrm{d}z^{\prime}\right)
\nonumber\\
&  \times\int\mathrm{d}p_{z}\int_{-\infty}^{0}\mathrm{d}z^{\prime}W\left(
\mathbf{r}^{\prime}+\frac{z^{\prime}}{p_{z}}\mathbf{p}+z^{\prime}%
\mathbf{e}_{z},\mathbf{p}+p_{z}\mathbf{e}_{z};0\right)
\end{align}
\end{widetext}
With this result it is easy to see that the stationary treatment of
decoherence in Sect.~\ref{sec:deco} is equivalent to the dynamic approach in
the present section. To facilitate the comparison we treat the present problem
with the method of Sect. \ref{sec:deco}. Take the beam to be in a nontrivial
stationary state at $z=0$,%
\begin{equation}
W_{\text{beam}}\left(  \mathbf{r},\mathbf{p}+p_{z}\mathbf{e}_{z}\right)
=g\left(  p_{z}\right)  w\left(  0;\mathbf{r,p}\right)  .
\end{equation}
In the case of coherent evolution (\ref{eq:freetrafo1}) the interference
pattern reads then%

\begin{equation}
Q\left(  z;\mathbf{r}\right)  \propto\int\mathrm{d}\mathbf{q~}e^{\mathrm{i}%
\mathbf{qr}/\hbar}\int\mathrm{d}p_{z}~\frac{p_{z}}{m_{\mathrm{p}}}g\left(
p_{z}\right)  \bar{w}\left(  z,p_{z};\mathbf{q}\right)  \label{Qstat}%
\end{equation}
with
\begin{align}
\bar{w}\left(  z,p_{z};\mathbf{q}\right)  :=&\frac{1}{\left(  2\pi\hbar\right)
^{2}}\int\mathrm{d}\mathbf{p}\mathrm{d}\mathbf{r~}\exp\left(  -\mathrm{i}%
\left(  \mathbf{r}+\frac{z}{p_{z}}\mathbf{p}\right)  \mathbf{q}/\hbar\right)
\nonumber\\
&\times w\left(  0;\mathbf{r},\mathbf{p}\right)  .
\end{align}
Using the same procedure as in Sect.~\ref{sec:deco} one finds how decoherence
events that take place at a constant rate $R\left(  z\right)  $ in
$(z;z+\mathrm{d}z)$ will modify the pattern (\ref{Qstat}). The result is given
by the expression in (\ref{Qstat}) if the $\bar{w}\left(  z;\mathbf{q}\right)
$ are replaced by
\begin{align}
\hat{w}\left(  z,p_{z};\mathbf{q}\right)  =&\exp\left(  -\int_{0}^{z}R\left(
z^{\prime}\right)  \left[  1-\eta\left(  \frac{z^{\prime}-z}{p_{z}}%
\mathbf{q}\right)  \right]  \mathrm{d}z^{\prime}\right) 
\nonumber\\
&\times \bar{w}\left(
z,p_{z};\mathbf{q}\right)  , \label{statdeco}%
\end{align}
where $\eta$ is the corresponding decoherence function. Clearly,
(\ref{statdeco}) is the analog of (\ref{finpara}) in the case of a stationary
description. The only difference is the appearance in (\ref{finpara}) of
$\bar{p}_{z}$ instead of $p_{z}$, which occurs because the additional
assumption of a strongly peaked longitudinal velocity distribution was
necessary in the time dependent calculation. The strong similarity between the
results (\ref{statdeco}) and (\ref{finpara}) shows that the treatment of
decoherence in Sect. \ref{sec:deco} is indeed equivalent to solving the master
equation in paraxial approximation.

\section{Conclusions}

\label{sec:conclusions} In this article we presented an analysis of Talbot-Lau
matter wave interference that provides a quantitative prediction of the
effects encountered in the experimental realization. It was shown that by
describing the stationary beam in terms of the Wigner function both the
interaction with the grating forces and the effects of markovian decoherence
can be incorporated analytically. In addition, the formulation allows one to
distinguish unambiguously the quantum phenomena from the effects of classical mechanics.

Recently, our theory was successfully applied to describe experiments with
large molecules
\cite{Brezger2002a,Hornberger2003a,Hackermuller2003a,Hackermuller2003b,Hackermuller2004a}%
. Correspondingly, the discussion of decoherence effects in the present
article was confined to those mechanisms most relevant in the interference of
fullerenes and biomolecules. Indeed, the interaction with gas molecules and
the emission of heat radiation are expected to be relevant sources of
decoherence for all large particles. Our formulation applies immediately to
those since the bulk properties of the particles, such as the polarizability
or the absorption cross section, were used to describe the environmental coupling.

Other decoherence effects might become relevant as the particles increase
further in complexity. In particular, those couplings that entangle the
center-of mass motion with the rotation of the particle or with its internal
degrees of freedom become sources of decoherence. In these cases the observed
loss of visibility, which is inevitable if the detection is insensitive to the
relative coordinates, can be calculated with the same approach as discussed above.

The grating interaction will also require a more refined treatment at some
point. The eikonal approximation ceases to be valid for particles of
increasing size, because they interact stronger and at the same time they will
have a longer interaction time. A more careful evaluation of the propagation
through the grating will be needed in those cases.

A final remark concerns the ease of incorporating decoherence effects in the
present formulation of matter wave interference. It draws heavily on the fact
that one is able to separate the rate of decoherence events from their effect.
It seems that this approach, that avoids the solution of a master equation in
time, can be a transparent way of treating markovian dynamics. It is
vindicated by the comparison with the more conventional solution of a
corresponding markovian master equation that was presented in the last section
of this article.

\section{Acknowledgments}

We acknowledge many helpful discussions with Bj\"{o}rn Brezger and Anton
Zeilinger. K.H. thanks Bassano Vacchini for discussions on collisional
decoherence. This work was supported by the Austrian FWF in the programs START
Y177 and SFB 1505 and by the Emmy-Noether program of the Deutsche Forschungsgemeinschaft.

\bibliographystyle{apsrev}

\end{document}